\documentclass{JFM-FLM_Au}
\usepackage{bm}

\lefttitle{H.-L. Wu, A. Xu and H.-D. Xi}
\righttitle{Journal of Fluid Mechanics}

\title{Super-resolution reconstruction of turbulent flows from a single Lagrangian trajectory}

\author{Hua-Lin Wu\aff{1}, Ao Xu\aff{1,2} \and Heng-Dong Xi\aff{1,2}}

\affiliation{
\aff{1}Institute of Extreme Mechanics, School of Aeronautics, Northwestern Polytechnical University, Xi'an 710072, PR China
\aff{2}National Key Laboratory of Aircraft Configuration Design, Key Laboratory for Extreme Mechanics of Aircraft of Ministry of Industry and Information Technology, Xi'an 710072, PR China
}

\corresau{Ao Xu, \email{axu@nwpu.edu.cn}}

\begin{document}
\maketitle

\begin{abstract}
We studied the reconstruction of turbulent flow fields from trajectory data recorded by actively migrating Lagrangian agents. 
We propose a deep-learning model, track-to-flow (T2F), which employs a vision transformer as the encoder to capture the spatiotemporal features of a single agent trajectory, 
and a convolutional neural network as the decoder to reconstruct the flow field. 
To enhance the physical consistency of the T2F model, we further incorporate a physics-informed loss function inspired by the framework of physics-informed neural network (PINN), 
yielding a variant model referred to as T2F+PINN. 
We first evaluate both models in a laminar cylinder wake flow at a Reynolds number of $Re = 800$ as a proof of concept. 
The results show that the T2F model achieves velocity reconstruction accuracy comparable to that of existing flow reconstruction methods, 
while the T2F+PINN model reduces the normalised error in vorticity reconstruction relative to the T2F model. 
We then apply the models in a turbulent Rayleigh--B\'enard convection at a Rayleigh number of $Ra = 10^{8}$ and a Prandtl number of $Pr = 0.71$. 
The results show that the T2F model accurately reconstructs both the velocity and temperature fields, 
whereas the T2F+PINN model further improves the reconstruction accuracy of gradient-related physical quantities, 
such as temperature gradients, vorticity and the $Q$ value, with a maximum improvement of approximately 60 \% compared to the T2F model. 
Overall, the T2F model is better suited for reconstructing primitive flow variables, while the T2F+PINN model provides advantages in reconstructing gradient-related quantities. 
Our models open a promising avenue for accurate flow reconstruction from a single Lagrangian trajectory.
\footnote{
This article may be downloaded for personal use only.
Any other use requires prior permission of the author and Cambridge University Press.
This article appeared in Wu \emph{et al.}, J. Fluid Mech. \textbf{1026}, A46 (2026) and may be found at \url{https://doi.org/10.1017/jfm.2025.11033}.
}
\end{abstract}

\begin{keywords}
  B\'enard convection, plumes/thermals, machine learning.
\end{keywords}


\section{Introduction}
\label{sec:Introduction}

Access to high-resolution spatiotemporal flow fields is critical for a wide range 
of real-world applications, including the autonomous navigation of aerial and underwater 
vehicles \citep{lawrance2009windpath, masmitja2023tracking, zhang2023auv},  
migration of microswimmers \citep{qiu2022microswimmer, qiu2022gyrotactic,
mousavi2024zooplankton, mousavi2025optimization} 
and environmental monitoring \citep{smith2021benthicrover}. 
For example, in unmanned aerial vehicles (UAVs) and underwater autonomous navigation, 
accurate knowledge of the underlying turbulent flow fields enables the implementation 
of globally optimal path planning algorithms including model predictive control 
\citep{krishna2022finite, krishna2023ftle} and adaptive control \citep{landau2011adaptive}, 
which can outperform local decision-making approaches such as reinforcement learning 
\citep{reddy2016soar, reddy2018glider, gunnarson2021navigation}. This capability 
allows autonomous vehicles to identify and exploit beneficial flow features (e.g. updrafts), 
thereby improving energy efficiency and extending operational endurance. 
However, in realistic atmospheric or ocean environments, direct measurements of the full Eulerian 
flow field are often infeasible due to limited sensor coverage and the high cost of deployment. 
Instead, the available observation data are typically a single Lagrangian trajectory, 
collected by mobile sensors mounted on the vehicles themselves. These measurements are 
inherently Lagrangian in nature and often represent the only accessible data under operational 
conditions \citep{calascibetta2023tracking, jiao2025sensing}. 
Several methods have been proposed to reconstruct Eulerian fields from Lagrangian observations. 
For example, FlowFit \citep{gesemann2016noisy} and VIC+ \citep{schneiders2016dense} 
reconstruct Eulerian fields using physics-constrained approaches that achieve accurate reconstructions when dense particle tracking data are available. 
However, in realistic scenarios of autonomous aerial or underwater navigation, only a single Lagrangian trajectory may be accessible, 
and the information contained in such sparse measurements is insufficient for these methods.
This situation poses a challenge: 
can we accurately reconstruct the flow field from a single Lagrangian trajectory?

This flow reconstruction challenge can be formulated as a super-resolution 
reconstruction problem, where the goal is to infer high-resolution flow fields from sparse 
and incomplete measurements. Conceptually, the task parallels classical image super-resolution 
in computer vision, where high-resolution images are reconstructed from their low-resolution 
counterparts \citep{wang2020srdl}. Given the sparsity of the available data, 
machine-learning-based super-resolution (MLSR) methods have emerged as promising tools 
to address this problem. Recent advances have extended MLSR methods to fluid flows by 
replacing the RGB (red, green and blue) image channels with physically meaningful 
quantities such as velocity or temperature fields \citep{fukami2023survey}. 
Building on this analogy, various machine learning architectures have been developed for 
flow-specific MLSR tasks. \cite{fukami2019superres} introduced a convolutional neural network (CNN) 
architecture for a two-dimensional laminar cylinder wake and homogeneous decaying turbulence. 
Subsequent extensions include a spatiotemporal MLSR method \citep{fukami2021spatiotemporal}, a Voronoi 
tessellation-assisted MLSR method \citep{fukami2021voronoi} and a single-snapshot 
MLSR method \citep{fukami2024snapshot}, each tailored to distinct application scenarios. 
The applicability of MLSR approaches across diverse flow configurations has also been 
demonstrated by \cite{liu2020deepturb}, \cite{nair2020rom}, \cite{zhou2022poreflow} and \cite{liu2026reconstructing}. 
To enhance the robustness and generalisability of these models, physics-informed loss functions that incorporate 
the governing equations of fluid dynamics have been introduced into the training process 
\citep{fukami2023survey}. These physical constraints may be imposed in 
unsupervised learning \citep{bode2021pigans, gao2021picnn} or incorporated 
as part of a hybrid loss function that combines physical consistency with a traditional mean 
square error loss function in supervised learning \citep{lee2019cylinder, ren2023physr}.
Recently, \cite{weiss2025temperature} demonstrated an elegant physics-based method for reconstructing the temperature field by solving a Poisson equation derived from applying the curl operator twice to the Navier--Stokes equations. 
Similar to MLSR methods, this temperature field reconstruction requires Eulerian measurements.

Despite recent advancements, reconstructing flow fields from sparse Lagrangian trajectory 
data remains more challenging than conventional super-resolution tasks. First, 
the input measurements consist of irregularly sampled and temporally evolving trajectories. 
This irregularity hinders effective feature extraction by conventional CNN-based MLSR methods, 
thereby motivating the development of alternative architectures capable of directly 
processing Lagrangian inputs \citep{fukami2021voronoi}.  
Second, in practical applications, trajectory data are often corrupted by sensor 
noise and localisation errors, which degrade signal quality and introduce 
uncertainties into the reconstructed flow fields. The reconstructions must remain 
physically consistent under such noisy conditions, particularly when gradient-related 
flow quantities (e.g. vorticity or velocity gradient) are of interest \citep{jiao2025sensing}. 
These quantities are highly sensitive to even minor spatial errors in the reconstructed 
primitive fields, and any lack of physical consistency may result in significant distortions 
of the underlying flow structures.

Together, these challenges underscore the need for machine learning models that not 
only accommodate irregular and noisy Lagrangian trajectory data but also enforce physical 
consistency throughout the reconstruction process. In this work, we present a deep-learning model, termed track-to-flow (T2F), for reconstructing flow fields from the 
Lagrangian trajectories of self-propelling agents. The T2F model integrates a 
vision transformer (ViT) to capture spatiotemporal patterns within the trajectory data, 
and a CNN as the decoder to generate flow fields in the 
vicinity of the agent trajectories. In addition, a physics-informed loss function is 
incorporated to enhance the physical consistency, particularly in gradient-related quantities. 
The rest of this paper is organised as follows. In \S~\ref{sec:Numerical methods}, 
we introduce the T2F model in detail. In \S~\ref{sec:cylinder}, we validate the model in a 
laminar cylinder wake flow, serving as a 
proof-of-concept test. In \S~\ref{sec:RB}, we extend the application to the turbulent 
Rayleigh--B\'{e}nard (RB) convection, a canonical flow system representative of convection 
in the atmosphere and oceans. The main findings of this work are summarised in \S~\ref{sec:conclusion}.

\section{Numerical methods}
\label{sec:Numerical methods}

\begin{figure}
  \centering
  \includegraphics[width=0.9\textwidth]{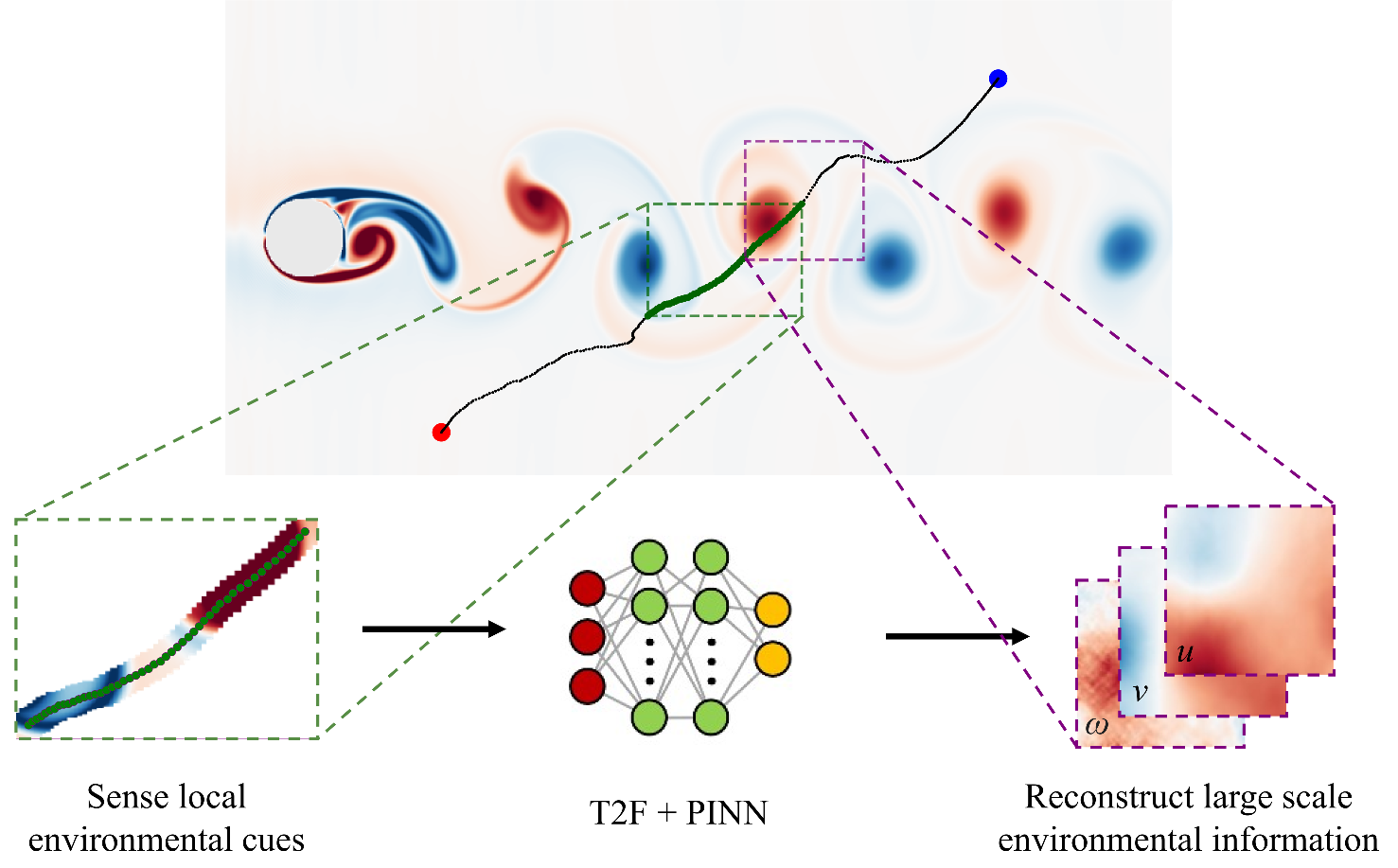}
  \caption{Overview of the T2F model for reconstructing flow fields in the cylinder wake.
  An actively navigating Lagrangian agent collects local flow cues along its trajectory,
  which are subsequently used to infer the surrounding Eulerian flow field.}
  \label{fig:framework}
\end{figure}

An overview of the T2F model is illustrated in figure \ref{fig:framework}. We first employ reinforcement learning 
to train self-propelling Lagrangian agents to perform point-to-point migration tasks within a 
flow environment, thereby generating agent trajectories as training data. Subsequently, the 
T2F model takes the trajectory information from the self-propelling agents as input and outputs 
the flow field in the vicinity of those trajectories.

\subsection{Migration of self-propelling agents}
\label{sec:Migration of self-propelling agents}

In this work, we consider an inertialess self-propelling agent model 
\citep{cichos2020active}, which is described as

\begin{equation}
\bm{u}_{\text{agent}} 
= \bm{u}_{\text{fluid}} + \bm{u}_{\text{propel}} 
= \bm{u}_{\text{fluid}} + U_{\text{propel}} 
[\cos(\theta), \sin(\theta)],
\end{equation}
\begin{equation}
\bm{x}_{\text{agent}}(t + dt) 
= \bm{x}_{\text{agent}}(t) + 
\bm{u}_{\text{agent}}(t) dt,
\end{equation}
where $dt$ is the time step.
Here, $\bm{u}_{\text{agent}}$ and $\bm{x}_{\text{agent}}$ denote 
the velocity and position of the agent, respectively;  
$\bm{u}_{\text{fluid}}$ is the local fluid velocity; and $U_{\text{propel}}$ 
is the self-propelling velocity magnitude generated by the agent. The agent moves at a constant 
speed $U_{\text{propel}}$ and directly controls its swimming direction $\theta$.
This is a toy model that describes the kinematics of UAVs in the atmosphere or 
autonomous underwater vehicles in the ocean. The model is justified by the fact that, 
in realistic atmospheric or oceanic scenarios, the characteristic length scale 
of the vehicles (of the order of metres) is several orders of magnitude smaller 
than that of the atmospheric or oceanic convection layer (typically kilometres). 
Similar dynamic models have been adopted in previous works 
\citep{biferale2019zermelo, borra2022reinforcement, krishna2022finite, monthiller2022surfing}.

To control the migration behaviour of a self-propelling agent within a flow environment, 
we employ reinforcement learning, a model-free control strategy rooted in behavioural 
psychology, in which an agent learns optimal actions through trial-and-error interactions 
with its environment \citep{sutton1998rl}. 
Reinforcement learning has been increasingly applied in fluid mechanics, 
including drag reduction \citep{wang2022active,zhou2025channel}, heat transfer enhancement \citep{zhou2025deep}, vortex shedding 
control \citep{li2022rlcylinder} and biologically inspired navigation tasks \citep{zhu2022predation}. In 
this work, we formulate a point-to-point migration problem, wherein agents are trained 
to reach randomly assigned target locations from randomly initialised starting points. The 
environmental cues available to the agent include its current position, its position 
relative to the target, the local fluid velocity and the target position. This information 
defines the observation state $s = [\bm{x}_{\text{agent}}, 
\Delta \bm{x}_{\text{agent}}, 
\bm{u}_{\mathrm{fluid}}, 
\bm{x}_{\mathrm{target}}]$, where 
$\Delta \bm{x}_{\text{agent}} = \bm{x}_{\text{target}} - \bm{x}_{\text{agent}}$. 
Based on this observation, the agent takes an action $a_t$, defined as the control 
of the propulsion direction $\theta$. The agent's behaviour 
is shaped by a reward function that encourages efficient navigation towards the target. 
Following \cite{gunnarson2021navigation}, we define the reward function as 

\begin{equation}
  r_t = -dt + 10\left[ 
  \frac{\|\bm{x}_{t-1} - \bm{x}_{\text{target}}\|}{U_{\text{propel}}} - 
  \frac{\|\bm{x}_t - \bm{x}_{\text{target}}\|}{U_{\text{propel}}} 
  \right] + r_{\text{bonus}},
\label{eq:reward}
\end{equation}
where

\begin{equation}
  r_{\mathrm{bonus}} = 
  \begin{cases}
  200, & \|\bm{x}_t - \bm{x}_{\text{target}}\| < H/36\\[6pt]
  0, & \text{others}
  \end{cases}
\end{equation}
Here, $\bm{x}_t$ and $\bm{x}_{t-1}$ denote the agent's position at the current and previous time steps, respectively,
and $H$ is the height of the computational domain.
The first term of \eqref{eq:reward} penalises time consumption, 
thereby encouraging the agent to navigate quickly. 
The second term of \eqref{eq:reward} rewards progress towards the target, 
while the last term of \eqref{eq:reward} provides a large terminal reward for successful 
arrival within a defined proximity to the target.

The reinforcement learning training is conducted using the soft actor-critic algorithm, 
which aims to maximise both the expected cumulative reward (i.e. successful task completion) 
and the entropy of the policy (i.e. encouraging exploration). 
The optimisation objective is defined as

\begin{equation}
  \pi^{*}(\theta)=\arg \max_{\pi} E_{\tau \sim \pi}\left[ \sum_{t=0}^{\infty}\left\{ r_t(s_{t},a_{t},s_{t+1})+\alpha H[\pi(\cdot | s_{t})] \right\} \right].
\end{equation}
Here, $\pi$ denotes the policy, 
represented by a neural network that maps the observation state $s_t$ 
to a Gaussian distribution over actions $a_t$. 
The notation $\pi(\cdot | s_t)$ denotes that the policy is 
stochastic. $\pi^*$ denotes the optimal policy, 
i.e. the policy with optimized parameters $\phi^*$. 
The trajectory $\tau = (s_0, a_0, s_1, a_1, \ldots, s_t, a_t)$ 
represents a sequence of states and actions generated 
by the policy, and $\tau \sim \pi$ indicates that the trajectory is sampled from $\pi$. 
The reward function is $r_t(s_t, a_t, s_{t+1})$ defined in \eqref{eq:reward}, 
and $H[\pi(\cdot | s_t)]$ is the entropy term that encourages exploration.

The entropy $H$ of the policy $\pi$ at state $s_t$ is computed as 

\begin{equation}
  H[\pi(\cdot|s_{t})]=E_{a_t \sim \pi(\cdot | s_{t})}[-\log \pi(a_t|s_{t})].
\end{equation}
For a Gaussian distribution $\pi(\cdot | s_{t})$ over actions $a_t$ with 
mean $\mu(s_{t})$ and standard deviation $\sigma(s_{t})$,
the entropy can be simplified as

\begin{equation}
  H[\pi(\cdot | s_{t})] = \frac{1}{2} \log\left( 2\pi e \sigma(s_{t})^2 \right).
\end{equation}
The entropy $H$ encourages exploration by favouring more stochastic policies. 
The parameter $\alpha$ is a trade-off coefficient that balances the reward and entropy terms. 
Further details of the soft actor-critic algorithm can be found in \cite{haarnoja2018sac}.

\subsection {Deep-learning model: track-to-flow}

\begin{figure}
  \centering
  \includegraphics[width=1.0\textwidth]{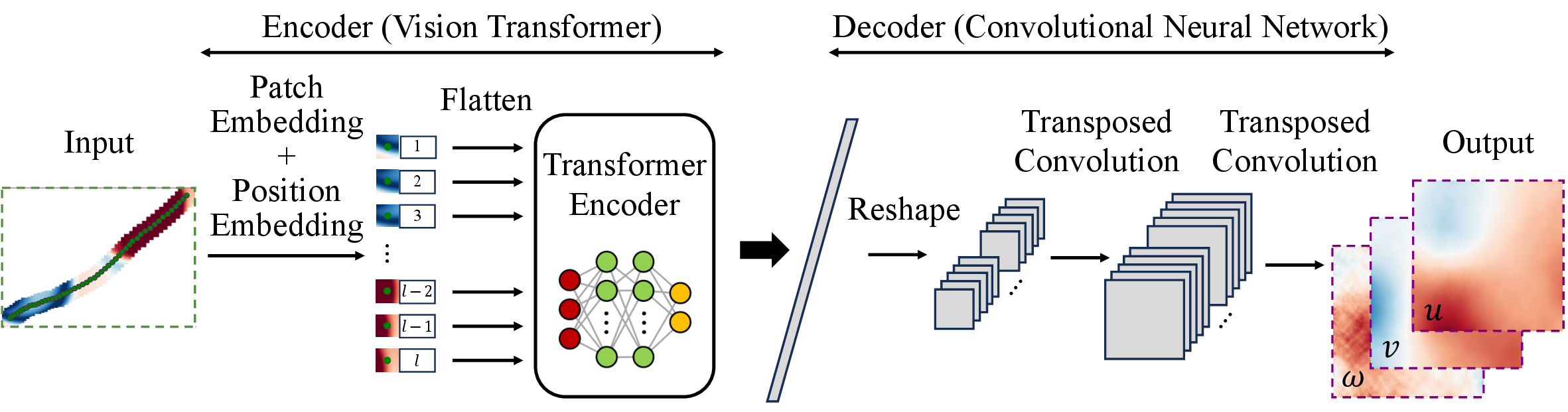}
  \caption{Schematic of the T2F model architecture. The model consists of a ViT encoder that extracts spatiotemporal features from Lagrangian
  trajectory data, followed by a CNN decoder that reconstructs the corresponding Eulerian flow field.}
  \label{fig_network}
\end{figure}

We develop the T2F deep-learning model to reconstruct flow fields from the Lagrangian 
trajectories of self-propelling agents. The T2F model adopts an encoder--decoder architecture 
comprising a ViT as the encoder and a CNN as the decoder (see figure \ref{fig_network}). Encoder--decoder architectures are widely employed 
in deep learning, particularly in natural language processing \citep{badrinarayanan2017segnet}
and computer vision \citep{cho2014nmt}. The encoder transforms the input sequence into a set 
of high-dimensional feature representations, which are subsequently utilised by the 
decoder to generate the desired output. Such architectures have been successfully applied 
to aerodynamic feature extraction under extreme conditions \citep{fukami2023grasping}.

In this study, the input to the model consists of localised trajectory data 
from self-propelling agents. These trajectories encode both fine-scale gradient 
information over short time scales and broader spatial--temporal correlations 
over longer horizons. Such multi-scale features are inherently difficult to extract 
using traditional methods. Interestingly, these input characteristics resemble 
those encountered in natural language processing and computer vision tasks, where 
close contextual relationships exist between adjacent words or pixels, while longer-range 
dependencies span across sentences or image regions. To this end, we adopt the 
Transformer architecture, which is capable of capturing both short- and long-range 
dependencies in sequential data. Specifically, we utilise the ViT as the encoder 
\citep{dosovitskiy2021vit}, as illustrated on the left-hand side of figure \ref{fig_network}. 
The ViT has demonstrated competitive performance in visual 
tasks by directly processing images as sequences of patches, which are small segments 
obtained by partitioning the input image. 
In the two-dimensional setting, 
we model a single particle trajectory as a short `video' composed of local flow snapshots.  
The input of the T2F model encoder is a four-dimensional tensor 
$x_0 \in \mathbb{R}^{\,l_p \times l_p \times l_t \times C}$,
where $l_p$ is the edge length of each square patch so that one patch spans 
an $l_p \times l_p$ neighbourhood of grid points; 
$l_t$ is the number of time steps sampled along the trajectory; 
and $C$ denotes the number of physical channels stored at every 
grid point (e.g. the velocity components ($u$,$v$), pressure, temperature, etc.).
Next, Transformer architecture is applied to extract 
spatiotemporal features from the Lagrangian input. 
The final output of the ViT encoder is a tensor $x_{\text{ViT}}\in\mathbb{R}^{l_t \times d_e}$, 
which represents a latent embedding of the input sequence.

Following feature extraction, a decoder is employed to reconstruct the corresponding 
Eulerian flow field. The decoder is based on a CNN, which is a class of deep-learning 
models widely used in image processing tasks \citep{li2021cnnreview}. Through multiple layers of 
convolution and pooling, CNNs progressively extract and refine hierarchical spatial 
features. In this study, we utilise the inverse operation of convolution, namely deconvolution, 
to reconstruct the flow field from the encoded features. Specifically, the deconvolution 
operation transforms an input tensor $x_{in} \in \mathbb{R}^{H_1\times W_1\times C_1}$ 
into an output tensor $ x_{out} \in \mathbb{R}^{H_2\times W_2\times C_2}$, as illustrated on the right-hand side of figure \ref{fig_network}. 
The encoded features are first reshaped into multi-channel matrices and then progressively 
upsampled through multiple deconvolution layers. The final output is the reconstructed flow field
$y \in \mathbb{R}^{H \times W \times C}$, 
where $C$ represents the number of physical quantities being reconstructed 
and $H \times W$ represents the spatial domain adjacent to the agent trajectories. 
Details of the T2F model architecture, including the number of layers and
hyperparameters, are provided in Appendix \ref{appArchitecture}.

\subsection {The physics-informed loss function}
\label{subsec:lossfunc}

We employ a physics-informed loss function inspired by the framework 
of physics-informed neural networks (PINNs), which are a class of mesh-free 
methods for solving partial differential equations using neural networks 
\citep{raissi2019pinn}. In conventional neural network training, data-driven 
models learn mappings between inputs and outputs by minimising a loss function 
defined over labelled datasets. 
The PINNs extend this paradigm by incorporating 
governing physical laws (typically represented as differential equations) directly 
into the loss function. This approach enables the neural network to learn solutions 
that approximately satisfy the underlying physics, even in the absence of 
dense or high-fidelity training data. Although PINNs offer significant advantages, 
they enforce physical constraints only approximately, treating the governing 
equations as soft constraints. As a result, their accuracy may degrade when 
solving forward problems at moderate-to-high Reynolds numbers \citep{chuang2022pinnreport}. 
Nevertheless, PINNs have demonstrated success in inverse problems, where system 
parameters or hidden fields must be inferred from sparse or noisy observations. 
Representative applications include the inference of structural properties, pressure 
and velocity fields \citep{raissi2019vortex,raissi2020hidden,boster2023aivelocimetry}, 
as well as the reconstruction of experimental flow velocity fields from noisy measurements 
\citep{cai2021espresso, kontogiannis2022jointrecon, toscano2025aivt}.
It is worth mentioning that the philosophy of physics-informed approaches has also been applied to operators by embedding partial differential equations into the loss functions, 
such as a physics-informed neural operator, as \cite{zhao2025lesnets} demonstrated in the novel application of LESnets (large-eddy simulation nets).

In the following, we first describe the mean-squared error loss function used in 
the standard T2F model, which does not incorporate any physics-based constraint. 
In the standard T2F model, the loss function \( L_{\text{MSE}} \) is defined as

\begin{equation}
  L_{\text{T2F}} = L_{\text{MSE}} = \frac{1}{N \times C} \sum_{j=1}^{C} \sum_{i=1}^{N} (y_{\text{rec}}^{(i,j)} - y_{\text{ref}}^{(i,j)})^2
\end{equation}
where \(N = H \times W\) is the total number of spatial grid points 
and \(C\) is the number of physical quantities being reconstructed. 
The terms \(y_{\text{rec}}^{(i,j)}\) and \(y_{\text{ref}}^{(i,j)}\) denote the 
reconstructed value and the reference values, respectively, at the \(i\)th grid 
point for the \(j\)th physical quantity. Minimising this loss encourages the 
model to align its predictions closely with the ground -truth data.

To incorporate physical constraints, we augment the loss function with a physics-informed term, yielding the T2F+PINN model, in which the total loss 
function comprises a data loss \(L_{\text{MSE}}\) and an equation-based loss \(L_{\text{PDE}}\). 
The equation loss is derived from the residual of the governing partial differential equations, expressed in general form as

\begin{equation}
  \frac{\partial u}{\partial t} + \mathcal{N}[u] = 0, \quad x \in \Omega, \, t \in [0,T],
\end{equation}
where \( u(x,t) \) is the latent solution field, 
\( \mathcal{N} \) is a nonlinear differential operator, \( \Omega \) 
is the spatial domain of the equation and \([0, T]\) is the time interval. 
The residual function is defined as

\begin{equation}
  f = \frac{\partial u}{\partial t} + \mathcal{N}[u],
\end{equation}
which quantifies the degree to which the reconstructed field violates the governing equations. 
The equation loss \( l_{\text{PDE}} \) for a single equation is given by

\begin{equation}
  l_{\text{PDE}} = \frac{1}{N} \sum_{i=1}^{N} \left| f(t_i, y_i) \right|^2,
\end{equation}
where $N$ is the number of grid points. 
Here, $f(t_i, y_i)$ denotes the residual evaluated at the $i$th grid point, 
where $t_i$ is the time and $y_i$ is the reconstructed field value at that point.
For systems governed by multiple equations, 
the total equation loss is a weighted sum of individual residuals:

\begin{equation}
  L_{\text{PDE}} = \sum_{k=1}^{N_f} w_k \, l_{\text{PDE}}^k = \sum_{k=1}^{N_f} w_k \, \frac{1}{N} \sum_{i=1}^{N} \left| f_k(t_i, y_i) \right|^2,
\end{equation}
where $N_f$ is the number of governing equations, 
$f_k$ denotes the residual for the $k$th equation and $w_k$ is the corresponding weight. 
In summary, the full loss function for the physics-augmented T2F+PINN model is

\begin{equation}
L_{\text{T2F+PINN}} = w_{\text{data}} L_{\text{MSE}} + \sum_{k=1}^{N_f} w_k \, l_{\text{PDE}}^k,
\end{equation}
where \( w_{\text{data}} \) and \( w_k \) control the relative contributions 
of data fidelity and physical consistency, respectively.

In this study, the inclusion of the physics-informed loss function transforms the reconstruction task into an inverse problem, in which the model aims to infer the latent Eulerian fields from observed Lagrangian trajectories. 
In contrast to conventional PINN formulations, the absolute spatial coordinates  $\bm{x}$ and time $t$ are not supplied as explicit inputs to the network; 
instead, the model processes local Eulerian patches extracted along the particle trajectory, while spatiotemporal context is introduced only through learnable positional embeddings.
As a result, we cannot apply automatic 
differentiation to compute temporal derivatives (e.g.\,\(\partial u/\partial t,\ \partial T/\partial t\)). 
Instead, these temporal derivatives are calculated using the reference velocity and 
temperature fields obtained from numerical simulations. After training, we assess the 
reconstruction performance of both models using the normalised \(L_2\) error, which provides 
a scale-invariant measure of accuracy. For a single reconstruction, 
the normalised \(L_2\) error \(\epsilon\) is defined as

\begin{equation}
\epsilon = \frac{\| y_{\text{rec}} - y_{\text{ref}} \|_2}{\| y_{\text{ref}} \|_2}
\end{equation}
where $y_{\mathrm{rec}}$ and $y_{\mathrm{ref}}$ denote the reconstructed and reference fields, respectively, and $\lVert \cdot \rVert_{2}$ is the Euclidean norm.
This metric enables consistent comparisons across different datasets and physical quantities.

\section {Flow field reconstruction in cylinder wake}
\label{sec:cylinder}

\subsection{Simulation settings}

We evaluate the T2F and T2F+PINN models in a two-dimensional cylinder wake flow as a 
proof-of-concept test. The governing equations for the incompressible flow around a circular cylinder are

\begin{equation}
\nabla \cdot \bm{u} = 0,
\end{equation}

\begin{equation}
\frac{\partial \bm{u}}{\partial t} + \bm{u} \cdot \nabla \bm{u} = -\frac{1}{\rho}\nabla p + \nu \nabla^{2}\bm{u},
\end{equation}
where $\bm{u} = (u, v)$ is the velocity field, $\rho$ is the density, $p$ is the pressure and $\nu$ is the kinematic viscosity. To non-dimensionalise the equations, we introduce the following scaling:

\[
\bm{x}^{\ast} = \frac{\bm{x}}{D}, 
\qquad
t^{\ast} = \frac{t\,U_{\infty}}{D}, 
\qquad
\bm{u}^{\ast} = \frac{\bm{u}}{U_{\infty}}, 
\qquad
p^{\ast} = \frac{p}{\rho\,U_{\infty}^{2}},
\]
where $U_{\infty}$ is the free-stream velocity and $D$ is the cylinder diameter. The dimensionless governing equations then become

\begin{equation}
\nabla \cdot \bm{u}^{\ast} = 0,
\end{equation}

\begin{equation}
\frac{\partial \bm{u}^{\ast}}{\partial t} 
+ \bm{u}^{\ast} \cdot \nabla \bm{u}^{\ast} 
= -\nabla p^{\ast} + \frac{1}{Re} \nabla^{2}\bm{u}^{\ast},
\end{equation}
where the Reynolds number is defined as

\begin{equation}
Re = \frac{U_{\infty} D}{\nu}.
\end{equation}

The computational domain is set to $[12D, 6D]$ and the mesh resolution is 
$1024 \times 512$. The cylinder is placed at the centre of the domain at 
coordinates $(D, 3D)$. 
The simulations are performed using the open-source lattice Boltzmann solver Palabos \citep{latt2021palabos} and are cross-validated with our in-house lattice Boltzmann solver \citep{XuLi2023,XuLi2024}. 
A Reynolds number of $Re = 800$ is chosen. 
The resulting vorticity field exhibits a well-defined K\'{a}rm\'{a}n vortex street (see figure \ref{fig_smart_particle_trajs}).

\subsection{Migration of self-propelling agents} \label{sec:MSPA}

\begin{figure}
  \centering
  \includegraphics[width=0.8\textwidth]{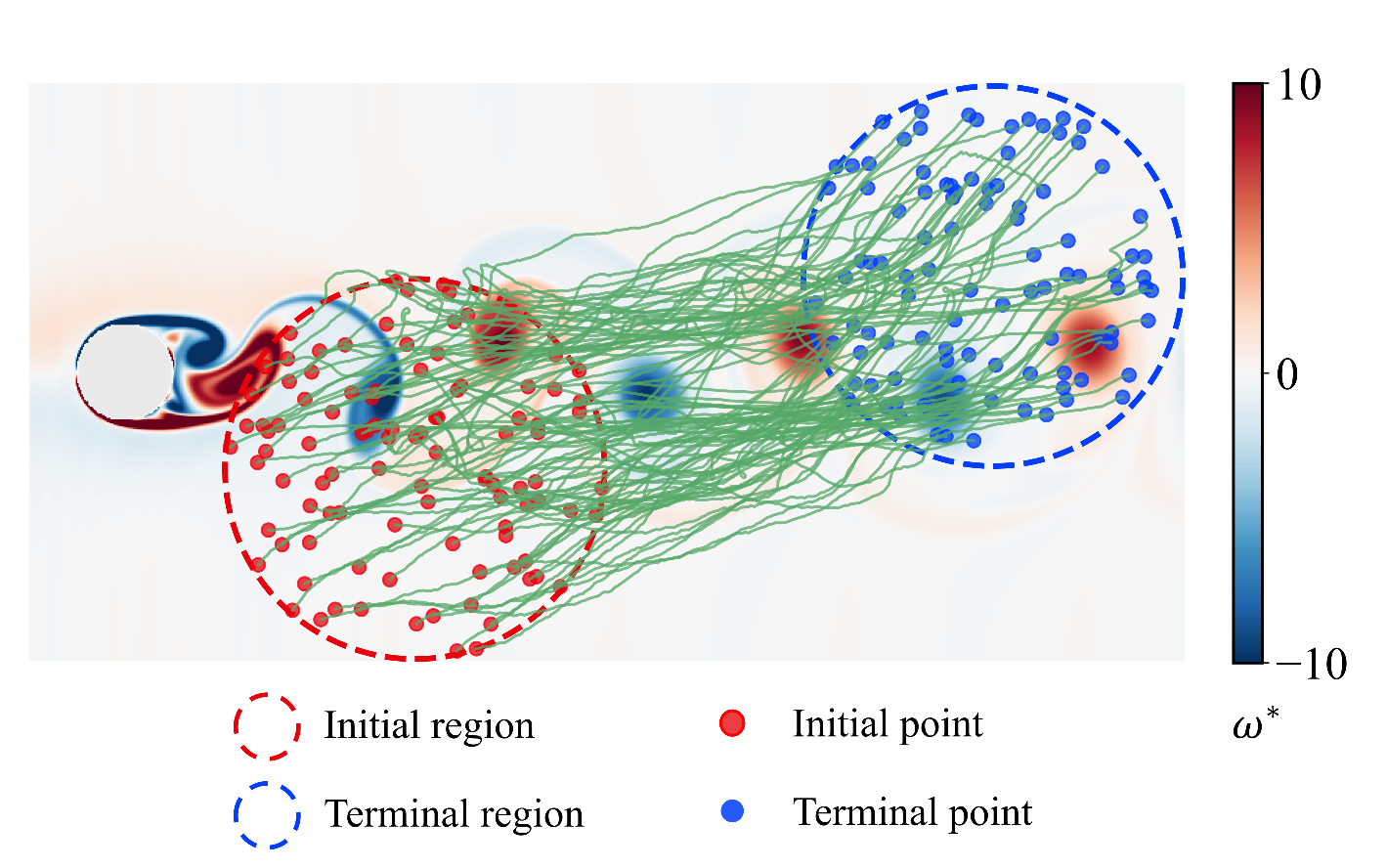}
  \caption{Trajectories of self-propelling agents navigating from the initial
  region (red) to the terminal region (blue) within the cylinder wake flow.
  The background contours represent the instantaneous out-of-plane vorticity field,
  illustrating the underlying flow structures guiding agent migration.}
  \label{fig_smart_particle_trajs}
\end{figure}

Using the simulated flow field, we train self-propelling agents via reinforcement learning 
to generate $n_{\text{traj}} = 100$ point-to-point migration trajectories. 
As a benchmark, we refer to the study by \cite{gunnarson2021navigation}. 
However, unlike their set-up, we modify the locations of the initial and terminal regions. 
In our configuration, agents migrate along the streamwise direction (see figure \ref{fig_smart_particle_trajs}), whereas \cite{gunnarson2021navigation} reported zigzag-like trajectories aligned with the spanwise direction. 
This modification slightly reduces task complexity while significantly improving training 
efficiency. In our case, the agent's reward signal saturates after approximately 1000 episodes 
and reaches its maximum value by around 4000 episodes. In contrast, the best-performing 
agent in the study of Gunnarson \emph{et al.} required roughly 5000 episodes to plateau and more than 
10 000 episodes to reach its optimal reward.

\subsection{Evaluation of the T2F and T2F+PINN models}

We evaluate the performance of the T2F and T2F+PINN models by reconstructing the velocity 
components in both the horizontal ($u_x$) and vertical ($u_y$) directions, corresponding to 
a total of $C = 2$ output channels. The reconstructed velocity fields are subsequently 
used to compute the out-of-plane vorticity, defined as $\omega_z = (\nabla \times \bm{u})_{z}$. 
The training dataset for both models is constructed as follows. First, the reinforcement 
learning agent described in \S~\ref{sec:MSPA} is used to generate $n_{\text{traj}} = 100$ 
point-to-point migration trajectories. From each trajectory, $n_{\text{sample}} = 10$ 
segments are extracted at randomly chosen initial times, each segment consisting of 50 
consecutive time steps. This yields a total of 
$n_{\text{train}} = n_{\text{traj}} \times n_{\text{sample}} = 1000$ training samples. 
The same procedure is applied in the testing phase to generate $n_{\text{test}} = 1000$ 
samples for evaluating the reconstruction accuracy of the models.

\begin{figure}
  \centering
  \includegraphics[width=0.8\textwidth]{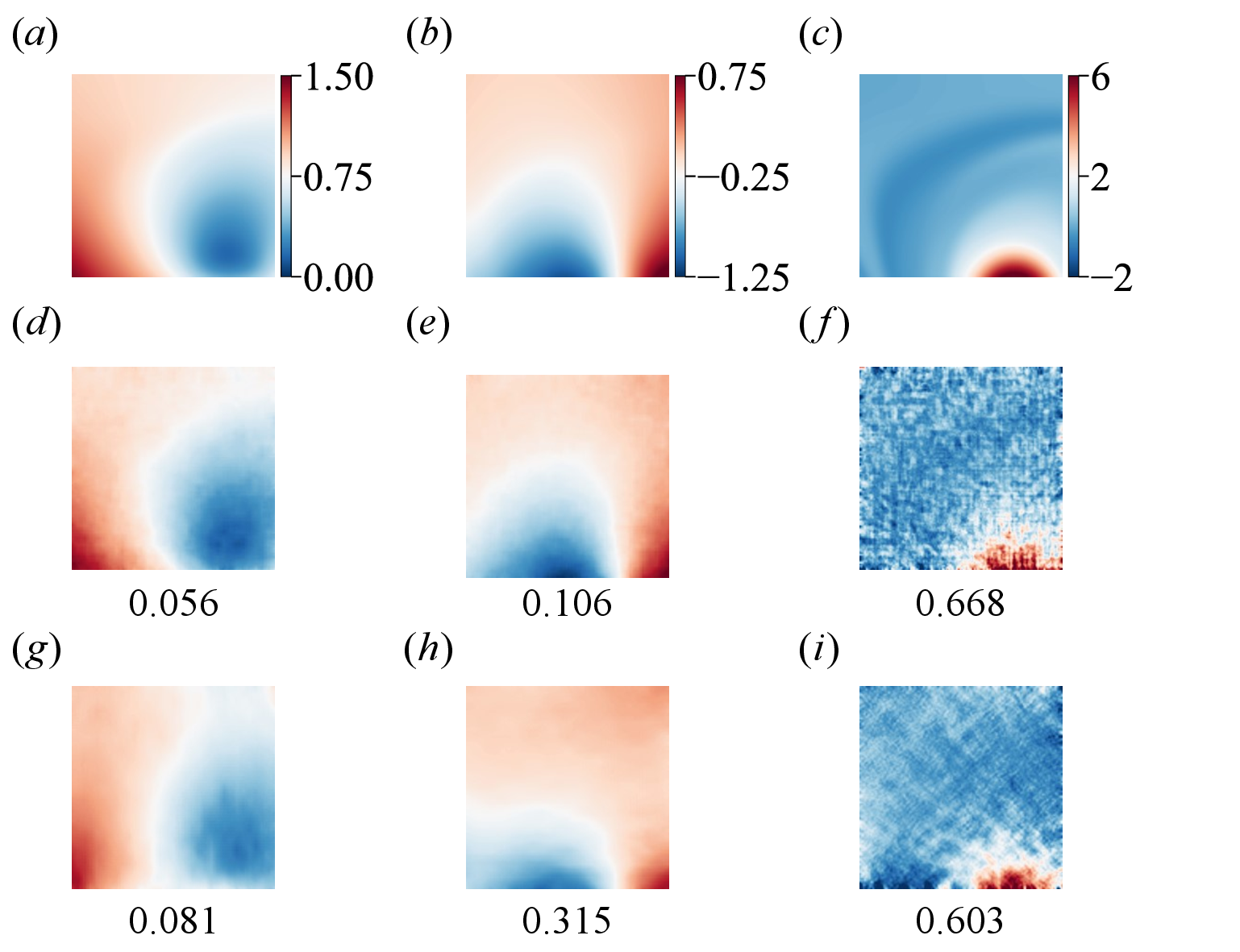}
  \caption{Reconstruction results from the T2F and T2F+PINN models for a representative input in the cylinder wake.
  Ground-truth fields of (\emph{a}) horizontal velocity $u_x^{\ast}$, (\emph{b}) vertical velocity $u_y^{\ast}$ and (\emph{c}) out-of-plane vorticity $\omega_z^{\ast}$.
  (\emph{d}--\emph{f}) Reconstructions by the T2F model.
  (\emph{g}--\emph{i}) Reconstructions by the T2F+PINN model.
  Listed values denote the normalised $L_2$ error $\epsilon$.}
  \label{fig_sample_comp_cylinder}
\end{figure}

Figure \ref{fig_sample_comp_cylinder} presents the reconstructed velocity fields obtained using the T2F and T2F+PINN models 
for a representative input, with the normalised $L_2$ error reported beneath each reconstructed 
flow field. We can see that both models are able to capture the spatial patterns of the flow structure. 
However, the reconstructions from the T2F model exhibit a noticeable blurring effect (see figure \ref{fig_sample_comp_cylinder}\textit{d,e}), 
resulting in the loss of fine-scale features. This blurring phenomenon is widely reported in 
flow-specific MLSR tasks across different flow scenarios 
\citep{fukami2019superres, zhou2022poreflow, liu2023multiresolution}. 
In contrast, the T2F+PINN model occasionally 
produces spatial misalignments in the reconstructed flow structures (see figure \ref{fig_sample_comp_cylinder}\textit{g,h}). 
This raises a natural question: how do such visual discrepancies in the primitive flow variables 
(e.g. velocity) affect the accuracy of gradient-based quantities (e.g. vorticity)?
When the reconstructed velocity field exhibits sharp but inconsistent transitions, as observed in the T2F model, 
the resulting vorticity computation becomes less accurate (see figure \ref{fig_sample_comp_cylinder}\textit{f}). This suggests that, 
although the purely data-driven T2F model can recover the overall flow structure, it lacks 
sufficient adherence to physical constraints necessary for accurately reconstructing gradient-based 
quantities. In contrast, the T2F+PINN model, by incorporating governing equations into the training 
process, effectively mitigates such inconsistencies and improves the accuracy of the reconstructed 
vorticity field (see figure \ref{fig_sample_comp_cylinder}\textit{i}). These differences underscore the importance of incorporating 
physics-informed constraints in enhancing the physical fidelity of reconstructions, particularly 
for gradient-related quantities.
To illustrate the dynamic reconstruction process of the T2F model for the cylinder wake flow, the corresponding video can be viewed in the supplementary movie 1 
available at \url{https://doi.org/10.1017/jfm.2025.11033}.

\begin{figure}
  \centering
  \includegraphics[width=0.8\textwidth]{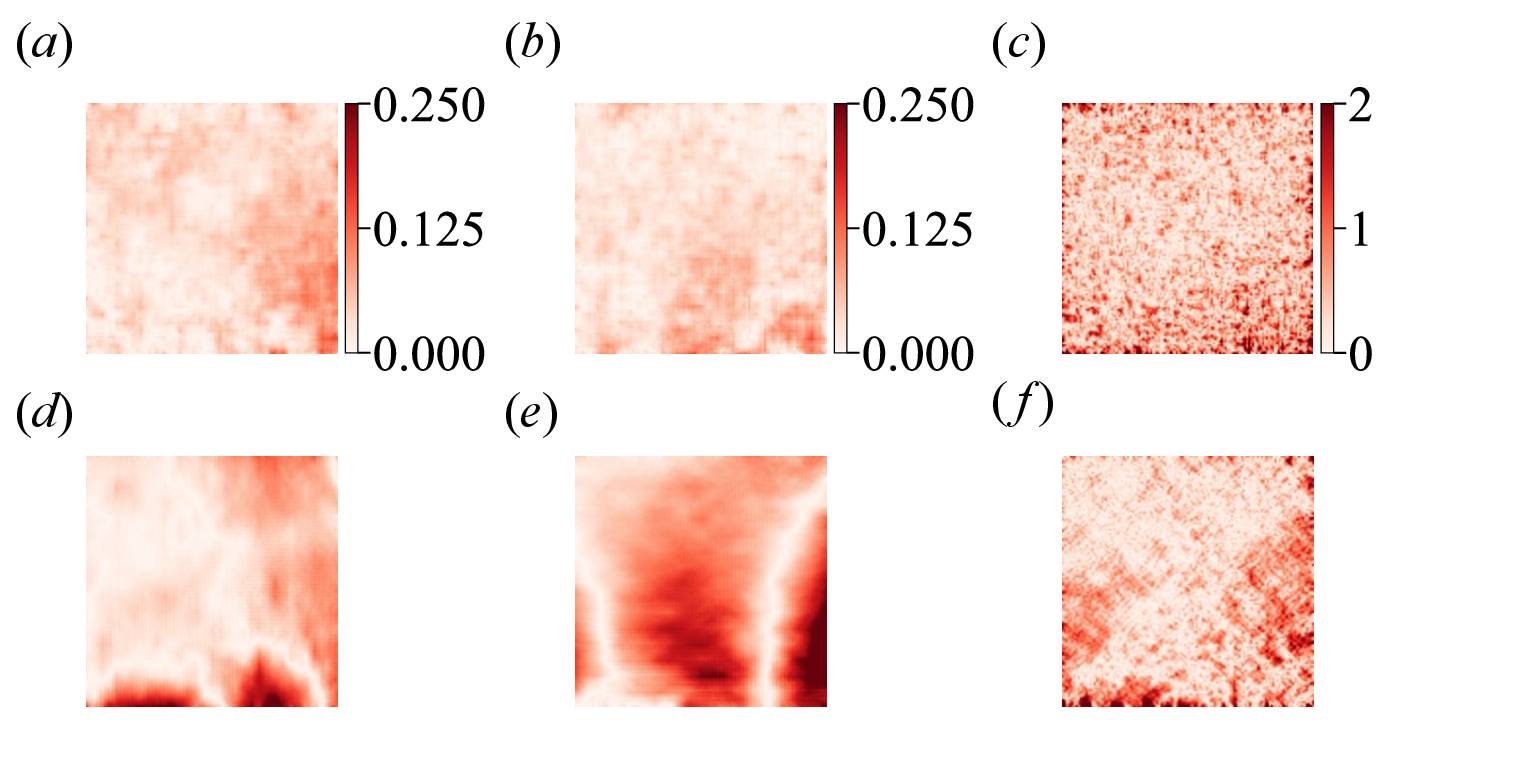}
  \caption{Pointwise reconstruction errors of the horizontal velocity $u_x^{\ast}$, vertical velocity $u_y^{\ast}$ and vorticity $\omega_z^{\ast}$ in the cylinder wake:
  (\emph{a}--\emph{c}) the T2F model, (\emph{d}--\emph{f}) the T2F+PINN model.}
  \label{fig_pointwise_error_cylinder}
\end{figure}

Figure \ref{fig_pointwise_error_cylinder} shows the pointwise error fields associated with the reconstructions produced 
by the T2F and T2F+PINN models. For the T2F model, the reconstruction errors appear to 
be randomly distributed across the domain, with no discernible spatial structure 
(see figure \ref{fig_pointwise_error_cylinder}\textit{a--c}). 
In contrast, the reconstruction errors from the T2F+PINN model 
exhibit geometrically structured patterns (see figure \ref{fig_pointwise_error_cylinder}\textit{d--f}), indicating that the error 
distribution is more closely aligned with the underlying physical processes. 
In particular, the reconstructed vortical structures in the T2F+PINN case display physically 
constrained translations and deformations, rather than spurious or uncorrelated distortions. 
Compared with the results in figure \ref{fig_sample_comp_cylinder}, these observations suggest that incorporating 
physics-informed constraints via PINN leads to more accurate reconstructions of vortical 
structures, thereby improving the recovery of gradient-based flow features.

\begin{figure}
  \centering
  \includegraphics[width=0.8\textwidth]{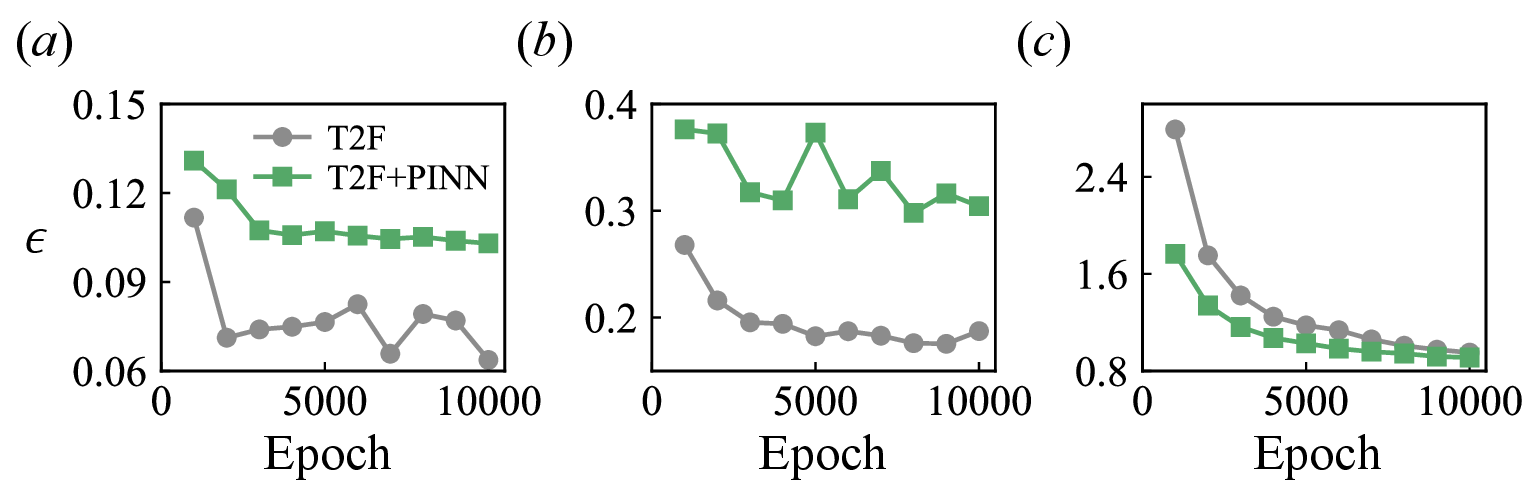}
  \caption{Evolution of the normalised $L_2$ errors over training epochs for the cylinder wake reconstruction task.
  Reconstruction errors for (\emph{a}) horizontal velocity $u_x^{\ast}$, (\emph{b}) vertical velocity $u_y^{\ast}$ and (\emph{c}) vorticity $\omega_z^{\ast}$, comparing the performance of the T2F and T2F+PINN models.}
  \label{fig_l2error_lineplot}
\end{figure}

We also evaluated reconstruction errors across all $n_{\text{test}} = 1000$ samples in the test set. 
Figure \ref{fig_l2error_lineplot} shows the averaged normalised $L_{2}$ error for the reconstructed variables as a 
function of training epoch. For the T2F model, the errors in the reconstructed velocity 
component $u_x^{\ast}$ and $u_y^{\ast}$ at $10\,000$ training epochs are approximately 0.06 and 0.19, 
respectively, which are much lower than those of the T2F+PINN model (see figure \ref{fig_l2error_lineplot}\textit{a,b}), 
corresponding to reductions of approximately 38.1 \% and 38.4 \%. In contrast, the vorticity error 
of the T2F+PINN model at $10\,000$ epochs is approximately 0.91, slightly lower than that of the 
T2F model, with a reduction of around 4.2 \% (see figure \ref{fig_l2error_lineplot}\textit{c}). These differences reflect the learning 
behaviours of the two models. The purely data-driven T2F model rapidly captures the dominant 
velocity structures but struggles to learn accurate velocity gradients, leading to higher errors 
in vorticity. The T2F+PINN model, on the other hand, incorporates physics-based constraints via 
the governing equations. While this results in a modest degradation in velocity reconstruction 
compared with the T2F model, it significantly enhances the fidelity of the reconstructed vorticity field. 
Similar trends have been reported by \cite{yousif2021hifrecon}, who observed that 
optimisation using only data loss produced distorted fluctuations in velocity components, 
whereas incorporating physics-based loss reduced sharpness in the flow-field details. 
We attribute this trade-off to the inherently multi-objective nature of the T2F+PINN framework. 
In addition to the data loss minimised by the single-objective T2F model, it introduces 
a physics-based loss term (see \S~\ref{subsec:lossfunc}). During training, these competing objectives 
require compromise, forcing the T2F+PINN model to balance physical consistency against direct 
data fidelity. As a result, when evaluated purely in terms of data loss (i.e.\ the $L_{2}$ 
error for primitive variables), the T2F+PINN model may underperform relative to the T2F model.
It is noteworthy that the fluctuations observed in figure \ref{fig_l2error_lineplot}(\textit{a},\textit{b}) reflect variations in test-set performance, rather than instability in the training process. 
Instead, training convergence is confirmed by monitoring the loss on a small held-out validation set during training. 
To demonstrate that the model has reached convergence, the corresponding training and validation loss curves are provided in Appendix \ref{appConvergence}.

To benchmark our method, we compare its performance with that of the MLSR approach by \cite{fukami2019superres}, 
which reconstructs high-resolution Eulerian fields from downsampled Eulerian inputs. 
Although the problem settings differ substantially, as our model infers flow fields from 
sparse Lagrangian trajectories, whereas \cite{fukami2019superres} reconstruct high-resolution fields from uniformly downsampled low-resolution data, 
we think such a comparison is still informative in interpreting the achievable reconstruction error levels. 
In the study by \cite{fukami2019superres}, a normalized $L_2$ error of approximately $\epsilon = 0.04$ was reported, when reconstructing a $192 \times 112$ high-resolution field from a $12 \times 7$ low-resolution input using 1000 snapshots. 
In comparison, our T2F model was also trained on 1000 samples, and it achieves a normalised $L_2$ error of approximately $\epsilon = 0.06$ when reconstructing a $128 \times 128$ velocity field from $2 \times 2 \times 50$ trajectory-based measurements. 
This similarity in error magnitude demonstrates the data efficiency of our proposed T2F model, particularly considering its reliance on sparse, irregular and non-grid-aligned Lagrangian inputs.

\begin{table}
  \centering
  \begin{tabular}{ccccccc}
   & \multicolumn{3}{c}{T2F model} 
   & \multicolumn{3}{c}{T2F+PINN model} \\ 
  Patch size $l_p$ 
   & $u_x^*$ & $u_y^*$ & $\omega_z^*$ 
   & $u_x^*$ & $u_y^*$ & $\omega_z^*$ \\
  1 & 0.080 & 0.213 & 1.084 & 0.116 & 0.306 & 0.933 \\
  2 & 0.081 & 0.223 & 1.012 & 0.114 & 0.309 & 0.920 \\
  4 & 0.080 & 0.217 & 0.883 & 0.114 & 0.298 & 0.893 \\
  8 & 0.080 & 0.210 & 0.825 & 0.113 & 0.284 & 0.866 \\
  \end{tabular}
  \caption{Normalised $L_2$ errors of the reconstructed velocity components and vorticity fields 
  in the cylinder wake for different patch sizes $l_p$. 
  The results are averaged over $n_{\text{test}}=1000$ test samples and six independent simulations.}
  \label{tab_lpatch_cylinder}
\end{table}

In this work, we model an unmanned aerial or underwater vehicle as a point-particle agent. 
In practice, however, such vehicles have finite size and often carry multiple sensors, 
making it reasonable to assume that local flow information in the vicinity of the particle is accessible. 
Accordingly, in our simulations each temporal slice of the trajectory corresponds not to 
a single-point measurement, but to a square spatial patch of size $l_p \times l_p$ 
centred on the particle position, containing $C$ physical variables ($C = 2$ for $u$, $v$). 
We investigate the robustness of the T2F and T2F+PINN models under varying
patch size $l_p$.
We consider four patch sizes: $l_p = 1$, $2$, $4$ and $8$ grid points. 
For $l_p = 1$, the input patch is a single point, which represents the limiting scenario in which the particle probes only a single point per time step, and it is obtained by bilinear interpolation of the nearest grid point. 
The results averaged over $n_{\text{test}}=1000$ test samples and six independent simulations for T2F and T2F+PINN model are shown in table \ref{tab_lpatch_cylinder}. 
For the cases $l_p=1$ and $l_p=2$ (corresponding to typical UAV scenarios with one or four sensors), 
the T2F model is more accurate for reconstructing the primary flow variables (e.g. velocity components $u_x^*$ and $u_y^*$), 
whereas the T2F+PINN model has an advantage in reconstructing gradient-related quantities (e.g. vorticity $\omega_z^*$). 
For larger patch sizes $l_p=4$ and $l_p=8$ (corresponding to UAV scenarios with 16 or 64 sensors), the T2F model outperforms the T2F+PINN model for all variables. 
We attribute this behaviour to the increased information content available by larger patch sizes, 
which reduces the relative impact of the PINN constraints.
Moreover, for both models, the reconstruction errors for $u_x^{\ast}$ and $u_y^{\ast}$
remain relatively stable across different patch sizes, with a maximum variation of about 5 \%. 
The vorticity error $\omega_z^{\ast}$ exhibits a slightly larger variations of about 14 \% 
when the patch size is increased from $l_p = 1$ to $l_p = 8$. 
Overall, the T2F and T2F+PINN models demonstrate robust performance across a range of patch sizes,
indicating their flexibility in handling different spatial scales of input data.

\begin{figure}
  \centering
  \includegraphics[width=0.8\textwidth]{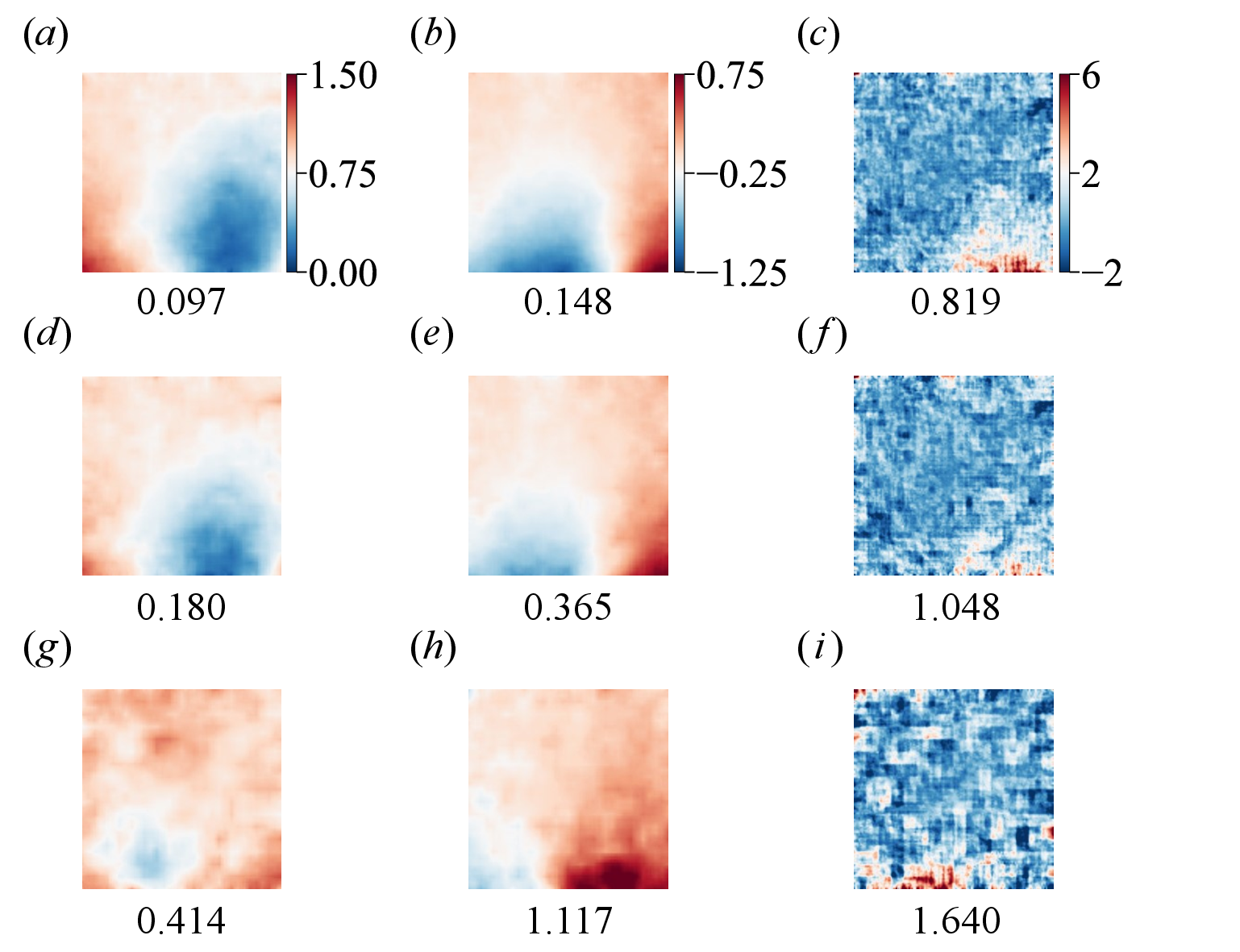}
  \caption{Reconstruction results of the T2F model under varying levels of input noise for a representative test sample:
  (\emph{a}--\emph{c}) horizontal velocity $u_x^{\ast}$, vertical velocity $u_y^{\ast}$ and vorticity $\omega_z^{\ast}$ at a noise level of $\eta = 0.1$;
  (\emph{d}--\emph{f}) reconstructions at $\eta = 0.2$;
  and (\emph{g}--\emph{i}) reconstructions at $\eta = 0.5$.
  Listed values denote the $L_2$ error $\epsilon$.}
  \label{fig_noise_naive}
\end{figure}

\begin{figure}
  \centering
  \includegraphics[width=0.8\textwidth]{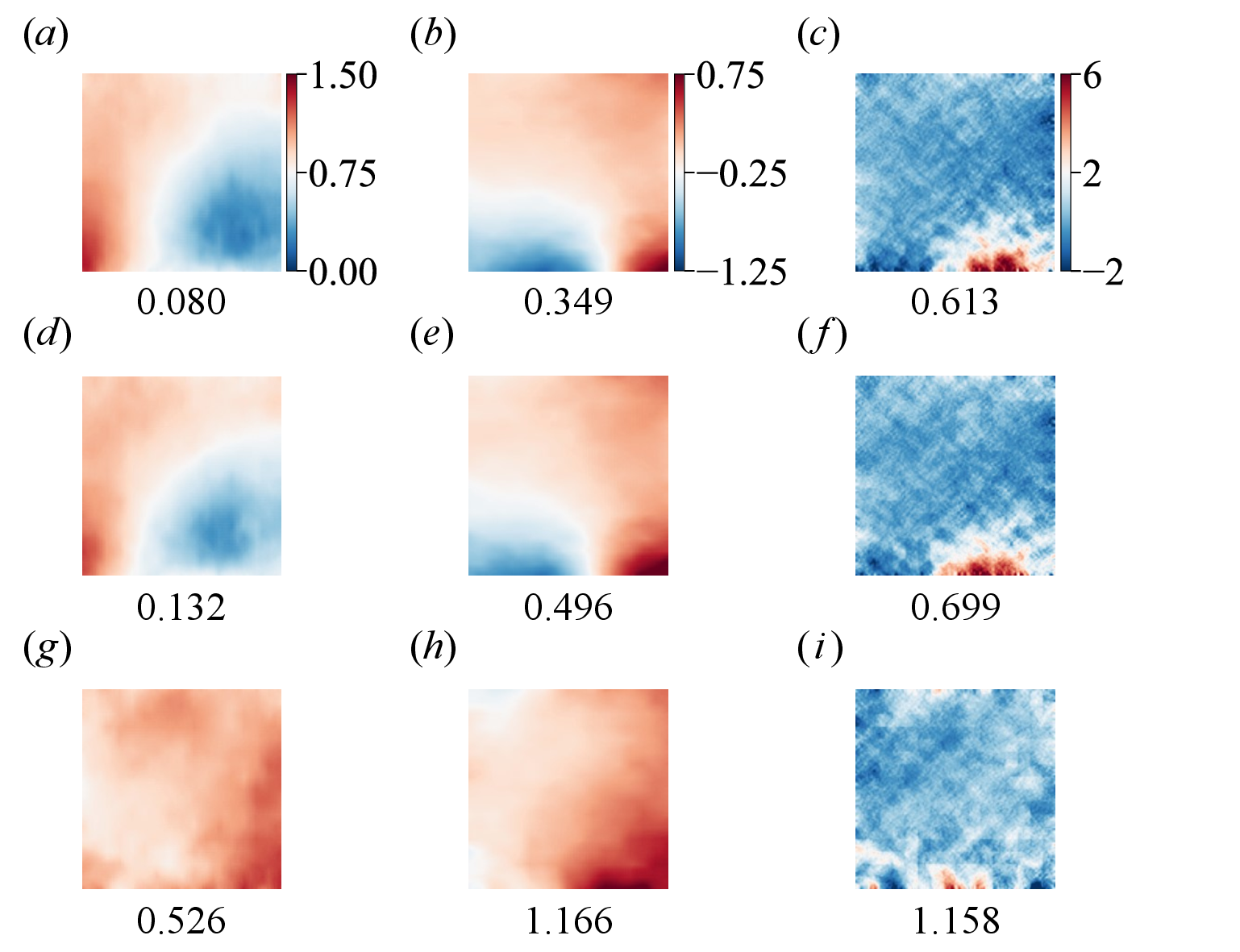}
  \caption{Reconstruction results of the T2F+PINN model under the same input noise levels as in figure~\ref{fig_noise_naive}:
  (\emph{a}--\emph{c}) horizontal velocity $u_x^{\ast}$, vertical velocity $u_y^{\ast}$ and vorticity $\omega_z^{\ast}$ at a noise level of $\eta = 0.1$;
  (\emph{d}--\emph{f}) reconstructions at $\eta = 0.2$;
  and (\emph{g}--\emph{i}) reconstructions at $\eta = 0.5$.
  Listed values denote the $L_2$ error $\epsilon$.}
  \label{fig_noise_pinn}
\end{figure}

In practical applications, observed environment cues may be contaminated by noise arising from sensor inaccuracies or environmental disturbances. 
To further evaluate the robustness of our model, 
we assess the performance of the T2F and T2F+PINN models under varying levels of input noise. 
Specifically, Gaussian noise is added to the input variable in the test set, 
which consists of the two velocity components $(u_x, u_y)$. 
For each component, random values are drawn independently from a normal distribution $\mathcal{N}(x_{\text{mean}}, x_{\max}/3)$, 
where $x_{\text{mean}}$ and $x_{\max}$ denote the mean and maximum values of that component, respectively. 
This design choice reflects realistic scenarios in autonomous aerial or underwater navigation, where onboard sensors measure local velocities, and adding noise to the measured variables therefore provides a faithful representation of sensor uncertainty. 
The noisy input is given by

\begin{equation}
x_{\text{noise}} = x_{\text{input}} + \eta \cdot \mathcal{N}(x_{\text{mean}}, x_{\max}/3),
\end{equation}
where $\eta$ controls the noise amplitude. We consider three noise levels 
of $\eta = 0.1$, $0.2$ and $0.5$. Figures \ref{fig_noise_naive} and \ref{fig_noise_pinn} present the reconstructed 
velocity components $u_x^{\ast}$, $u_y^{\ast}$ and the vorticity $\omega_z^{\ast}$ 
obtained from the T2F and T2F+PINN models, respectively, for a representative 
test case under each noise level. At the low noise level ($\eta = 0.1$), 
the T2F+PINN model achieves lower reconstruction errors in $u_x^{\ast}$ (0.080 versus 0.097) 
and $\omega_z^{\ast}$ (0.613 versus 0.819), while yielding a slightly higher 
error in $u_y^{\ast}$ compared with the T2F model (see figures \ref{fig_noise_naive}\textit{a--c} and \ref{fig_noise_pinn}\textit{a--c}). 
As the noise level increases to $\eta = 0.2$, the T2F model experiences substantial 
degradation in velocity reconstruction, with relative error increases 
of 85.6 \% for $u_x^{\ast}$ and 146 \% for $u_y^{\ast}$. In contrast, the T2F+PINN model 
exhibits improved robustness, with smaller error increases of 65.0 \% and 42.1 \% 
for $u_x^{\ast}$ and $u_y^{\ast}$, respectively. For the vorticity field, the T2F+PINN 
model shows only a 14.0 \% increase in error, compared with a 28.0 \% increase for the 
T2F model (see figures \ref{fig_noise_naive}\textit{d--f} and \ref{fig_noise_pinn}\textit{d--f}). Under high-noise levels ($\eta = 0.5$), 
both models fail to reconstruct coherent vortex structures, with large errors across 
all fields. Nevertheless, the T2F+PINN model continues to yield relatively lower 
errors in $\omega_z^{\ast}$, reflecting its superior resilience to noise 
(see figures \ref{fig_noise_naive}\textit{g--i} and \ref{fig_noise_pinn}\textit{g--i}). In summary, as noise levels increase, the T2F model 
exhibits a significant decrease in reconstruction accuracy, especially for vorticity, 
whereas the T2F+PINN model maintains more stable performance across a wide range of 
noise levels.

\begin{figure}
  \centering
  \includegraphics[width=0.8\textwidth]{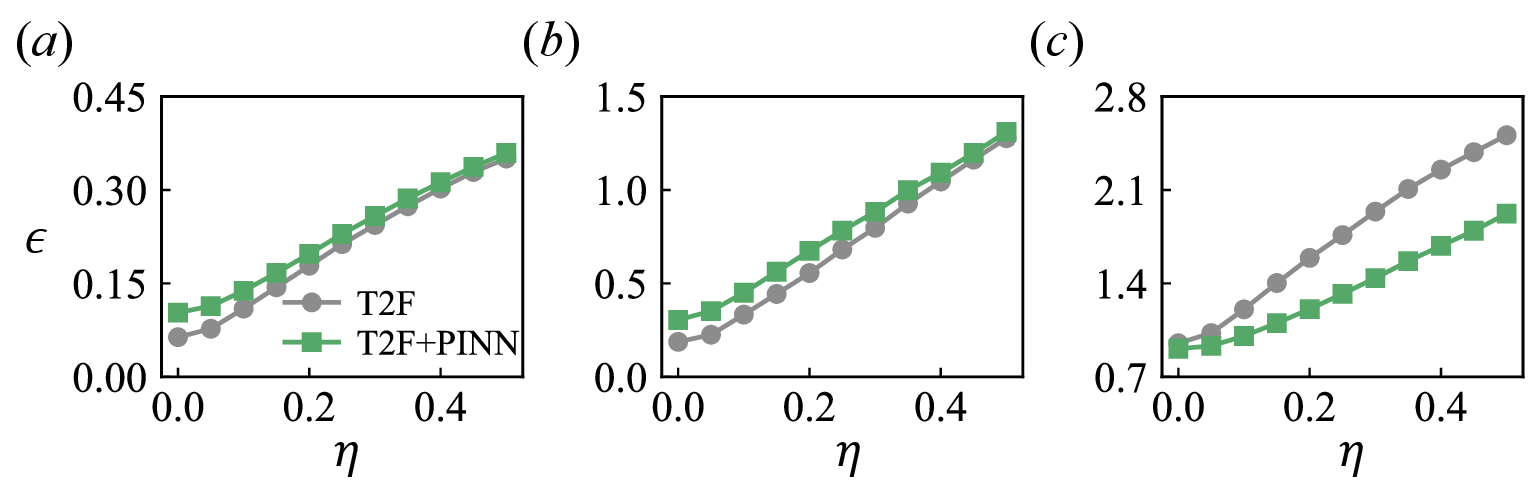}
  \caption{Normalised $L_2$ errors $\epsilon$ of the reconstructed (\emph{a}) $u_x^{\ast}$, (\emph{b}) $u_y^{\ast}$ and (\emph{c}) $\omega_z^{\ast}$ as functions of input noise
  levels $\eta$, for the T2F and T2F+PINN models in the cylinder wake.}
  \label{fig_noise_lineplot}
\end{figure}

We further assess the influence of input noise on the T2F and T2F+PINN models by computing 
the average reconstruction error over the entire test set consisting of $n_{\text{test}} = 1000$ 
samples. Figure \ref{fig_noise_lineplot} presents the variation of the normalised $L_{2}$ error $\epsilon$ with 
respect to the noise amplitude $\eta$ for the velocity components $u_x^{\ast}$ and $u_y^{\ast}$ 
as well as the vorticity $\omega_z^{\ast}$. As shown in figure \ref{fig_noise_lineplot}(\textit{a},\textit{b}), the T2F model 
exhibits slightly lower reconstruction errors for $u_x^{\ast}$ and $u_y^{\ast}$ at low noise 
levels ($\eta < 0.1$). However, its performance degrades more rapidly as noise increases, 
resulting in error levels comparable to those of the T2F+PINN model at higher noise ($\eta = 0.5$). 
In contrast, the T2F+PINN model displays a more gradual increase in error, demonstrating 
enhanced robustness in reconstructing primitive variables under noisy input conditions. 
For the vorticity field (figure \ref{fig_noise_lineplot}\textit{c}), the T2F+PINN model consistently outperforms the T2F 
model across all noise levels. Its error increases at a slower rate, indicating that the 
incorporation of physics-based constraints via the PINN framework effectively mitigates 
the degradation in gradient-based quantities caused by input noise. These findings confirm 
that although both models are affected by noise in the input trajectories, the T2F+PINN model 
exhibits superior robustness, particularly in reconstructing derived flow features such 
as vorticity. Similar robustness of physics-informed loss function has also been demonstrated 
in classical flow-specific MLSR tasks \citep{fukami2023survey}.

\section {Flow field reconstruction in Rayleigh--B\'{e}nard convection}
\label{sec:RB}

\subsection{Simulation settings}

The RB convection is a canonical system for modelling buoyancy-driven flows in the atmosphere 
and oceans \citep{lohse2010smallscale,chilla2012rbnew,xia2013trends,
wang2020vibration,lohse2023ultimate3,lohse2024ultimate2,xia2023coherent,xia2025some}. In RB convection, 
thermal plumes emerge from the thermal boundary layers near the hot and cold walls and 
subsequently interact to form a coherent large-scale circulation structure. 
We simulate RB convection under the Oberbeck--Boussinesq approximation, wherein 
temperature is treated as an active scalar that modulates the velocity field via a 
buoyancy force. The governing equations for the RB convection system are given by

\begin{equation}
\nabla \cdot \bm{u} = 0,
\end{equation}

\begin{equation}
\frac{\partial \bm{u}}{\partial t} + \bm{u} \cdot \nabla \bm{u}
= -\frac{1}{\rho_0}\nabla P + \nu \nabla^{2}\bm{u}
+ g\beta_T(T - T_0)\hat{\bm{y}},
\end{equation}

\begin{equation}
\frac{\partial T}{\partial t} + \bm{u} \cdot \nabla T
= \alpha_T \nabla^{2} T,
\end{equation}
where $\bm{u} = (u, v)$, $P$ and $T$ denote the velocity, pressure and 
temperature fields, respectively; $\rho_0$ and $T_0$ are the reference density 
and temperature, respectively; $\hat{\bm{y}}$ is the unit vector in the direction 
of gravity; $g$ is the gravitational acceleration; and $\nu$, $\beta_T$ and $\alpha_T$ represent 
the kinematic viscosity, thermal expansion coefficient and thermal diffusivity, 
respectively. With the following scaling:

\begin{equation}
\bm{x}^{\ast} = \frac{\bm{x}}{H}, 
\qquad
t^{\ast} = \frac{t}{\sqrt{H/\bigl(g\beta_{T}\Delta_{T}\bigr)}}, 
\qquad
\bm{u}^{\ast} = \frac{\bm{u}}{\sqrt{g\beta_{T}H\Delta_{T}}},
\end{equation}

\begin{equation}
P^{\ast} = \frac{P}{\rho_{0} g \beta_{T}\Delta_{T} H},
\qquad
T^{\ast} = \frac{T - T_{0}}{\Delta_{T}},
\end{equation}

the governing equations can then be rewritten in dimensionless form as

\begin{equation}
\nabla \cdot \bm{u}^{\ast} = 0,
\end{equation}

\begin{equation}
\frac{\partial \bm{u}^{\ast}}{\partial t^{\ast}}
+ \bm{u}^{\ast} \cdot \nabla \bm{u}^{\ast}
= -\nabla P^{\ast}
+ \sqrt{\frac{Pr}{Ra}}\nabla^{2}\bm{u}^{\ast}
+ T^{\ast}\hat{\bm{y}},
\end{equation}

\begin{equation}
\frac{\partial T^{\ast}}{\partial t^{\ast}}
+ \bm{u}^{\ast} \cdot \nabla T^{\ast}
= \sqrt{\frac{1}{Pr\,Ra}}\nabla^{2}T^{\ast}.
\end{equation}
Here $H$ is the cell height and it is chosen as the characteristic length; 
$t_{f} = \sqrt{H/(g\beta_{T}\Delta_{T})}$ is the free-fall time and it is chosen as the characteristic time; 
$T_{0}$ is the temperature of the cooling walls; and $\Delta_{T}$ is the temperature 
difference between the heating and cooling walls. 
The system is characterised by two dimensionless numbers, the Rayleigh number ($Ra$) and 
the Prandtl number ($Pr$), defined as

\begin{equation}
Ra = \frac{g\beta_{T}\Delta_{T}H^{3}}{\nu\alpha_{T}},
\qquad
Pr = \frac{\nu}{\alpha_{T}}.
\end{equation}

\begin{figure}
  \centering
  \includegraphics[width=0.8\textwidth]{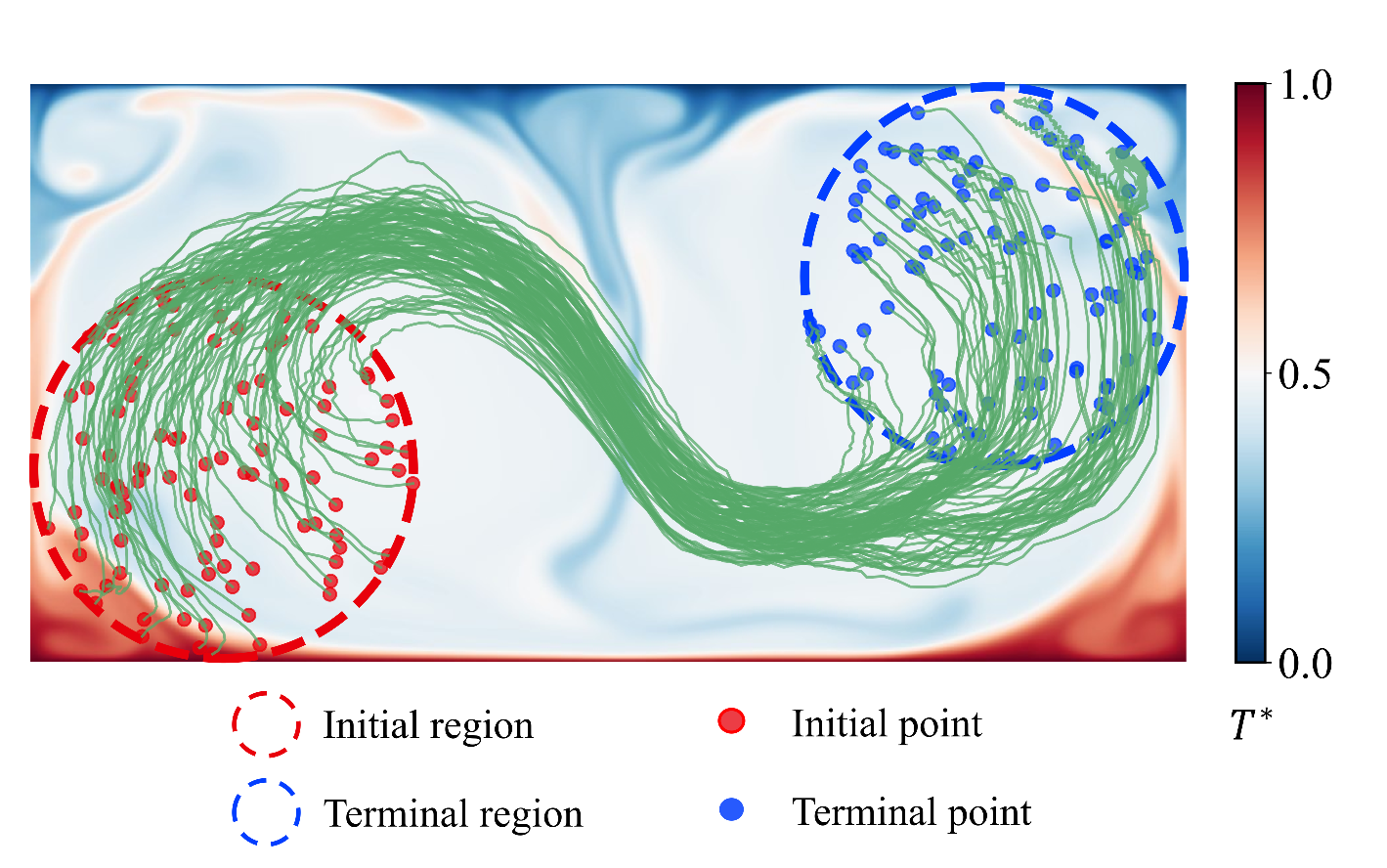}
  \caption{Trajectories of agents navigating from the initial region (red) to the
  terminal region (blue) in RB convection. The background contours show the
  instantaneous temperature field, highlighting thermal plumes and large-scale
  circulation structures.}
  \label{fig_smart_particle_trajs_RB}
\end{figure}

We employ the finite-volume solver OpenFOAM \citep{Weller1998} and cross-validate the results against our in-house lattice Boltzmann solver \citep{XuLi2023,XuLi2024} for RB convection at Rayleigh number $Ra = 10^{8}$ and Prandtl number $Pr = 0.71$. 
The computational 
domain is set to $[2H, H]$, corresponding to an aspect ratio $\Gamma = 2$. The domain is 
discretised using a uniform grid with a resolution of $1024 \times 512$. 
The simulation was carried out for a total of 200 free-fall times with an adaptive 
time step ensuring the Courant-Friedrichs-Lewy (CFL) number $\leq 0.3$. To eliminate transient effects, 
the initial 167 free-fall times were discarded as spin-up. 
Subsequently, snapshots were recorded every 0.01 free-fall times.
For training and testing, we used 1000 frames, corresponding to the statistically stationary interval between 167 and 177 free-fall times.
A representative snapshot of the temperature field is shown in figure \ref{fig_smart_particle_trajs_RB}, where thermal plumes emerging 
from the top and bottom boundaries are clearly visible. These plumes self-organize into 
two oppositely rotating large-scale circulations, characteristic of RB convection at this parameter regime 
\citep{lohse2023ultimate3, lohse2024ultimate2, shishkina2024ultimate1}.

\subsection{Migration of self-propelling agents}

Using the simulated flow field, we trained the self-propelling agents via reinforcement 
learning to generate $n_{\text{traj}} = 100$ trajectories, following the same navigation 
protocol as described in the cylinder wake case (see \S~\ref{sec:cylinder}). The resulting 
trajectories are shown in figure \ref{fig_smart_particle_trajs_RB}. Agents are initialised in a designated region near 
the lower-left corner of the domain. They are initialised, advected upward by the ascending 
hot plumes, traverse the bulk of the convection cell and subsequently descend along the 
cold plumes near the cell centre, eventually accumulating in a terminal region near the 
upper-right corner. These trajectory patterns are consistent with previous findings in a 
$\Gamma = 2$ convection system \citep{xu2022migration}, which highlight the 
agents' ability to actively exploit thermal structures in the environment for efficient 
navigation \citep{akos2010soaring,shepard2025might}.

\subsection{Evaluation of the T2F and T2F+PINN models}

Inspired by previous studies on navigation in fluid environments, which highlight the importance 
of velocity \citep{gunnarson2021navigation, jiao2025sensing} 
and temperature fields \citep{xu2022migration, xu2023longdistance}, 
we focus on reconstructing the horizontal and vertical 
velocity components $u_x$, $u_y$ and the temperature field $T$. These three variables are 
treated as separate channels in the model's input and output, corresponding to $C = 3$. 
From the reconstructed velocity and temperature fields, we further compute gradient-related quantities, 
including the out-of-plane vorticity $\omega_z = \nabla \times \bm{u}$, the horizontal 
temperature gradient $\partial_x T$ and the $Q$ value defined as $Q = \left( \lVert \boldsymbol{\Omega} \rVert^{2} - \lVert \bm{S} \rVert^{2} \right) /2$.
Here, $\boldsymbol{\Omega} = [\nabla \bm{u} - (\nabla \bm{u})^{\mathrm{T}}]/2$ is the 
antisymmetric vorticity tensor and $\bm{S} = [\nabla \bm{u} + (\nabla \bm{u})^{\mathrm{T}}]/2$ 
is the symmetric strain-rate tensor. 
The training and testing configurations follow those used in the cylinder wake reconstruction (see \S~\ref{sec:cylinder}). Specifically, we generate 
$n_{\text{traj}} = 100$ agent trajectories via point-to-point migration, and extract 
$n_{\text{train}} = 1000$ training samples and $n_{\text{test}} = 1000$ test samples to 
evaluate model performance.

\begin{figure}
  \centering
  \includegraphics[width=0.8\textwidth]{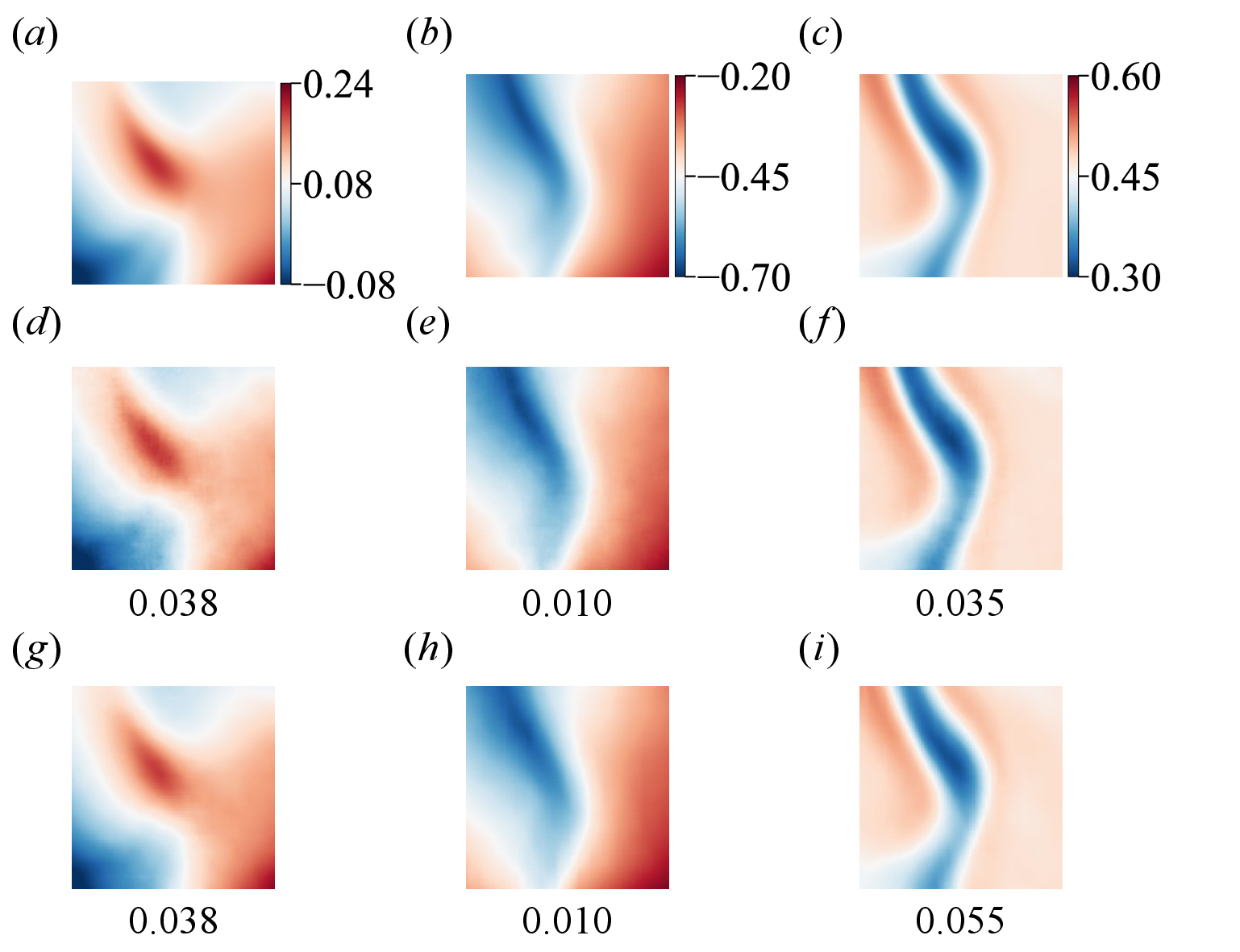}
  \caption{Reconstruction results of the T2F and T2F+PINN models for a representative input in RB convection.
  Ground-truth fields of (\emph{a}) horizontal velocity $u_x^{\ast}$, (\emph{b}) vertical velocity $u_y^{\ast}$ and (\emph{c}) temperature $T^{\ast}$.
  (\emph{d}--\emph{f}) Reconstructions by the T2F model.
  (\emph{g}--\emph{i}) Reconstructions by the T2F+PINN model.
  Listed values denote the normalised $L_2$ error $\epsilon$.}
  \label{fig_uvt_comp}
\end{figure}

\begin{figure}
  \centering
  \includegraphics[width=0.8\textwidth]{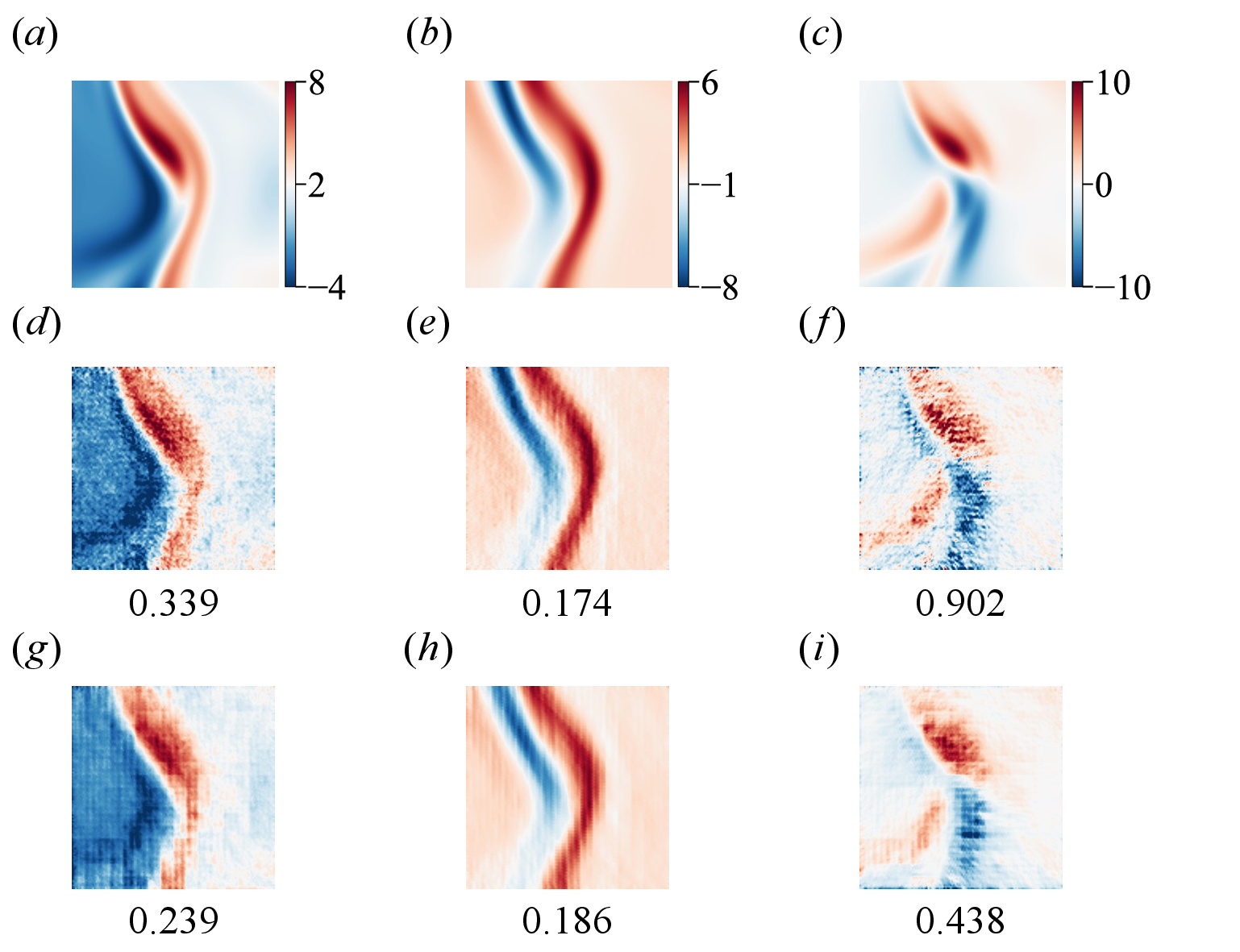}
  \caption{Reconstruction results of gradient-based quantities in RB convection.
  Ground-truth fields of (\emph{a}) out-of-plane vorticity $\omega_z^{\ast}$, (\emph{b}) the horizontal temperature gradient $\partial T^{\ast} / \partial x^{\ast}$ and (\emph{c}) the $Q$ value $Q^{\ast}$.
  (\emph{d}--\emph{f}) Reconstructions by the T2F model.
  (\emph{g}--\emph{i}) Reconstructions by the T2F+PINN model.
  Listed values denote the normalised $L_2$ error $\epsilon$.}
  \label{fig_grad_comp}
\end{figure}

Figure \ref{fig_uvt_comp} presents reconstruction results for the primitive flow variables, 
including the velocity components $u_x^{\ast}$, $u_y^{\ast}$ and the temperature 
field $T^{\ast}$, while figure \ref{fig_grad_comp} shows the reconstruction of gradient-related quantities, 
including the vorticity $\omega_z^{\ast}$, horizontal temperature gradient 
$\partial_{x^{\ast}}T^{\ast}$ and the $Q$ value $Q^{\ast}$, for a representative input 
sample using both the T2F and T2F+PINN models in the RB convection system. The numerical 
values shown beneath each reconstructed field indicate the corresponding normalized 
$L_{2}$ error $\epsilon$. We can see that both models successfully capture the 
spatial flow structures. Specifically, the differences in reconstruction errors 
of $u_x^{\ast}$ and $u_y^{\ast}$ between the two models are within 1 \% (see figure \ref{fig_uvt_comp}\textit{d,e,g,h}). 
However, the temperature field reconstructed by the T2F+PINN model exhibits an 
error approximately 57 \% higher than that of the T2F model (see figure \ref{fig_uvt_comp}\textit{f,i}), 
suggesting that the T2F model is more effective at reconstructing 
primitive physical variables in this case. Nevertheless, the T2F model displays noticeable 
blurring in the reconstructed velocity and temperature fields, which leads to 
degraded accuracy in the derived gradient-related quantities (see figure \ref{fig_grad_comp}\textit{d--f}). 
In contrast, the T2F+PINN model mitigates such artefacts by incorporating physical 
constraints from the governing equations during training, resulting in flow 
reconstructions that are more physically consistent with the underlying dynamics 
(see figure \ref{fig_grad_comp}\textit{g--i}). The representative sample captures the cold plume located 
near the centre of the RB convection domain, which is an important feature for 
flow perception and environment inference by navigating agents \citep{xu2023longdistance}. It is 
worth noting that in this specific case, the T2F model achieves a slightly lower error 
in the reconstructed temperature gradient. This discrepancy is attributed to 
stochastic variability within the test dataset rather than indicating a 
consistent performance trend.
To illustrate the dynamic reconstruction process of the T2F model for the RB convection, the corresponding video can be viewed in the supplementary movie 2.

\begin{figure}
  \centering
  \includegraphics[width=0.8\textwidth]{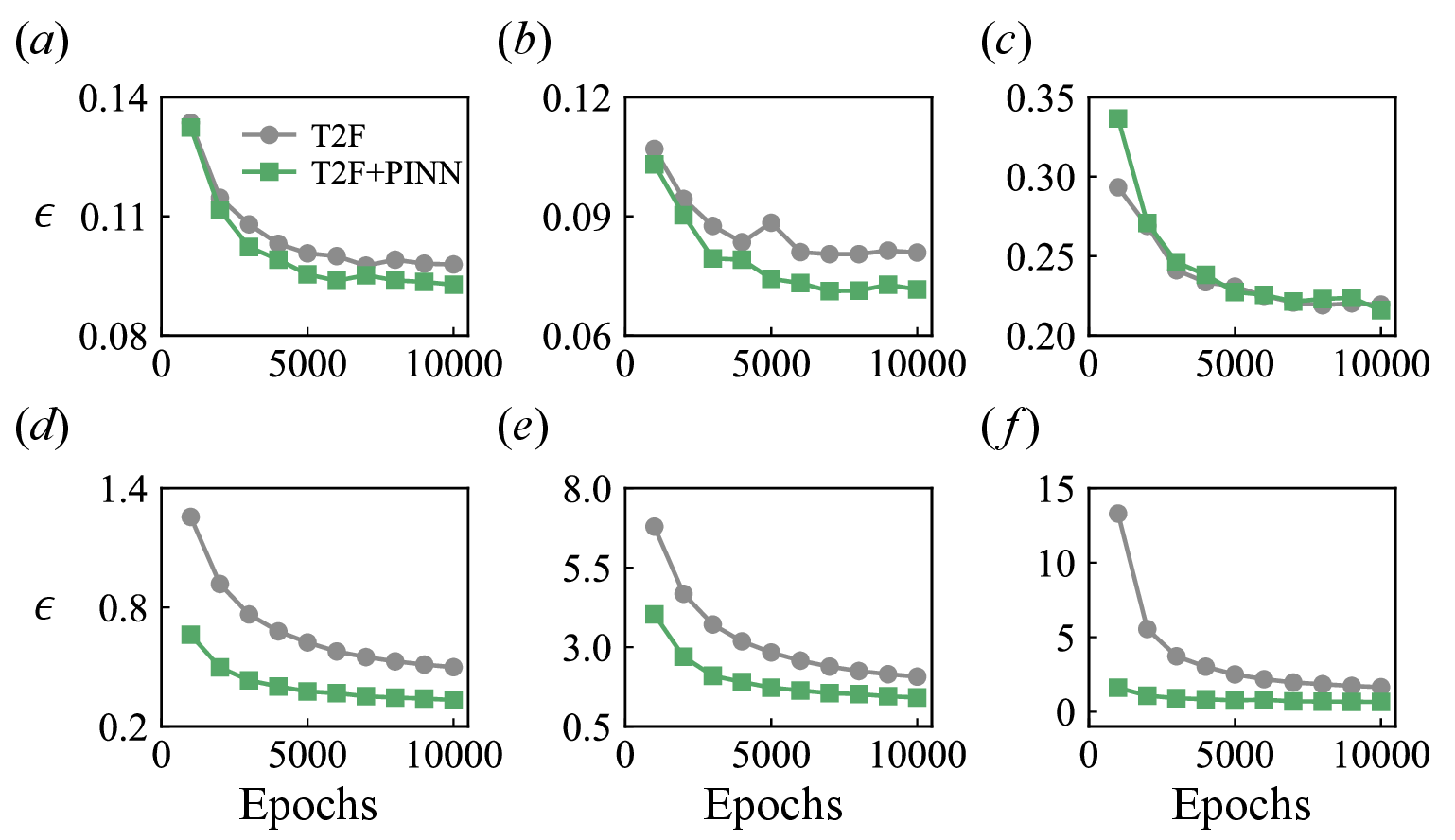}
  \caption{Evolution of the normalised $L_2$ errors over training epochs for the T2F and T2F+PINN models in RB convection.
  (\emph{a}--\emph{c}) The primitive variables $u_x^{\ast}$, $u_y^{\ast}$ and $T^{\ast}$.
  (\emph{d}--\emph{f}) The gradient-based quantities $\omega_z^{\ast}$, $\partial T^{\ast} / \partial x^{\ast}$ and $Q^{\ast}$.
  Results are compared between the T2F and T2F+PINN models.}
  \label{fig_l2error_lineplot_RB}
\end{figure}

Figure~ \ref{fig_l2error_lineplot_RB} presents the evolution of the 
normalised $L_{2}$ error with respect to training 
epochs for both the T2F and T2F+PINN models, evaluated over all $n_{\text{test}} = 1000$ 
samples in the test set. We consider both the primitive variables $u_x^{\ast}$, $u_y^{\ast}$ and $T^{\ast}$ (see figure \ref{fig_l2error_lineplot_RB}\textit{a--c}), as well as the gradient-related quantities including 
the vorticity $\omega_z^{\ast}$, the horizontal temperature gradient $\partial_{x^{\ast}}T^{\ast}$ 
and the $Q$ value $Q^{\ast}$ (see figure \ref{fig_l2error_lineplot_RB}\textit{d--f}). 
For the primitive variables, both models exhibit a decreasing trend in reconstruction error 
as training progresses. The T2F model reaches a final 
normalised $L_{2}$ error of approximately 0.09 for $u_x^{\ast}$, 0.08 for $u_y^{\ast}$ and 0.21 for $T^{\ast}$ at 10 000 epochs. The corresponding values for the T2F+PINN model are 
0.09, 0.07 and 0.21, respectively. For the gradient-related quantities, 
the T2F+PINN model consistently outperforms the T2F model. 
At 10 000 training epochs, the T2F+PINN model achieves 
relative reductions in normalized $L_{2}$ error of 33.3 \% for $\omega_z^{\ast}$, 
31.6 \% for $\partial_{x^{\ast}}T^{\ast}$ and a substantial 60.1 \% for $Q^{\ast}$. 
These quantities involve spatial derivatives and are therefore more sensitive to local 
field smoothness and physical consistency, which are better preserved by the 
physics-informed constraints embedded in the T2F+PINN model. 
These results show that in the RB convection system, the T2F+PINN model excels in reconstructing
gradient-based quantities, while the purely data-driven T2F model 
shows minor advantages in reconstructing primitive variables. 
To demonstrate that the model has reached convergence, 
we also provide the corresponding training and validation loss curves in Appendix \ref{appConvergence}.
Those conclusions obtained for the $\Gamma = 2$ RB convection system are expected to 
generalise to larger $\Gamma$ systems; where the number of convection rolls 
increase with $\Gamma$ \citep{wang2020multiple}, and the agents migrate a longer 
distance in the horizontal direction to mimic the behaviour of long-distance-migrating 
birds or patrolling UAVs \citep{xu2023longdistance}.

\begin{figure}
  \centering
  \includegraphics[width=0.8\textwidth]{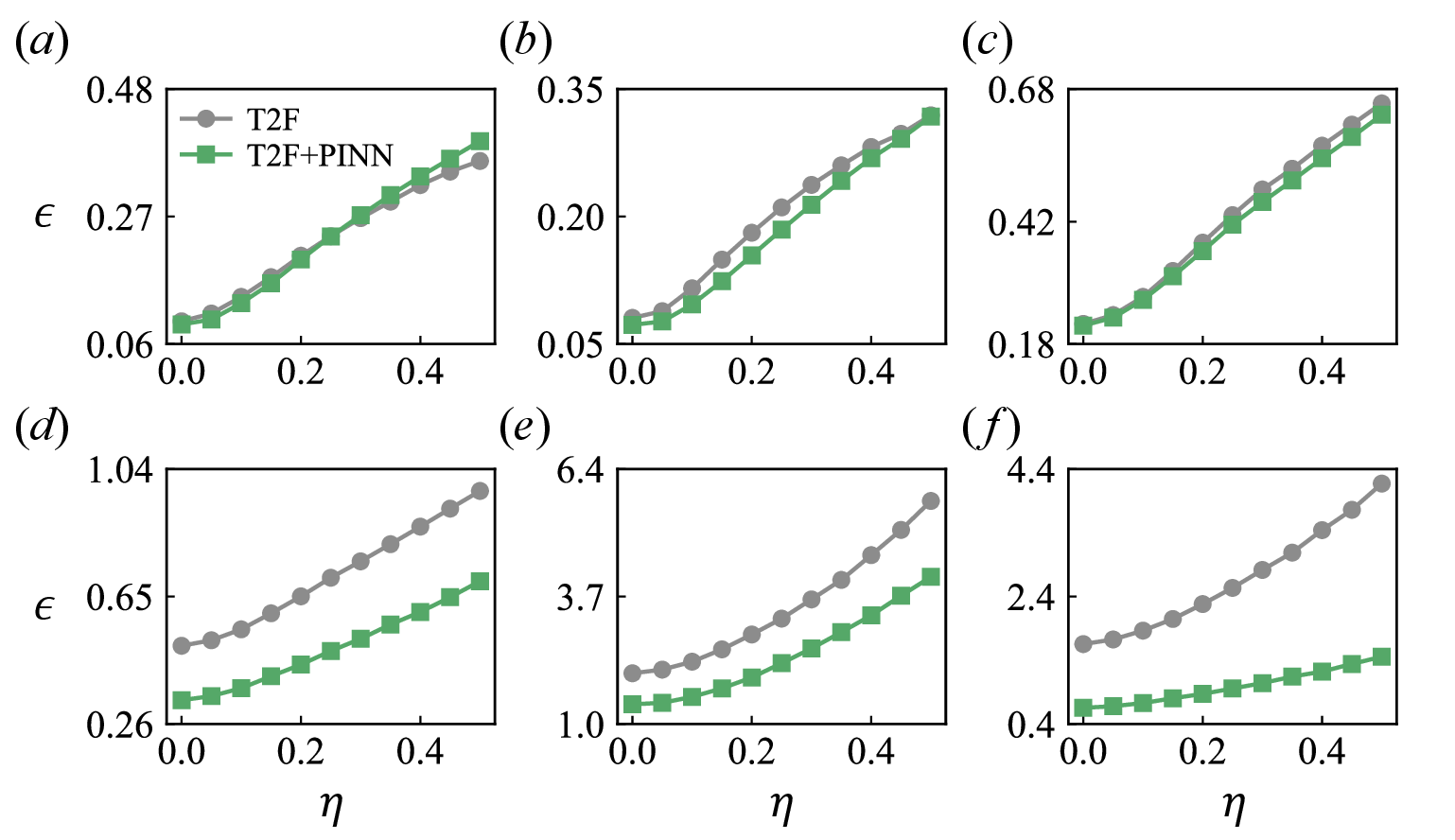}
  \caption{Normalised $L_2$ errors $\epsilon$ of the reconstructed primitive variables (\emph{a}) $u_x^{\ast}$, (\emph{b}) $u_y^{\ast}$ and (\emph{c}) $T^{\ast}$, and gradient-based quantities
  (\emph{d}) $\omega_z^{\ast}$, (\emph{e}) $\partial T^{\ast} / \partial x^{\ast}$ and (\emph{f}) $Q^{\ast}$ as functions of input noise
  levels $\eta$, for the T2F and T2F+PINN models in RB convection.}
  \label{fig_noise_lineplot_RB}
\end{figure}

We also test the influence of input noise on the T2F and T2F+PINN models by computing the average reconstruction error over the entire test set,
consisting of $n_{\text{test}} = 1000$ samples in RB convection. 
Noise is added to the input variable in the test set, which consists of the velocity components $(u_x, u_y)$ and temperature $T$, 
following the procedure described in \S~\ref{sec:cylinder}. 
Figure \ref{fig_noise_lineplot_RB} shows the variation of the normalised $L_{2}$ error $\epsilon$ 
with respect to the noise amplitude $\eta$ for primitive variables $u_x^{\ast}$, $u_y^{\ast}$, $T^{\ast}$, and for the gradient-based quantities $\omega_z^{\ast}$, $\partial T^{\ast} / \partial x^{\ast}$, $Q^{\ast}$. 
As shown in figure \ref{fig_noise_lineplot_RB}(\textit{a--c}), 
both models exhibit a gradual increase in error for the primitive variables as the noise level rises, 
and their performance remains broadly comparable across the full range of $\eta$. 
In contrast, for the gradient-based quantities (see figure \ref{fig_noise_lineplot_RB}\textit{d--f}), 
the T2F+PINN model consistently outperforms the T2F model, with its error increasing at a slower rate. 
These results are consistent with the findings in the cylinder wake case (see figure \ref{fig_noise_lineplot}), 
where the T2F+PINN model demonstrates enhanced robustness in reconstructing gradient-based quantities under noisy input conditions.

Next, we investigate the robustness of the T2F and T2F+PINN models under varying the 
patch size $l_p$ for the RB convection case. 
The results for the T2F and T2F+PINN models for $l_p = 1$, $2$, $4$ and $8$ 
grid points are shown in table \ref{tab_lpatch_RB}. 
Similar to the result in the cylinder wake, 
the reconstruction errors for both models
remain relatively stable across different patch sizes. 
The T2F model exhibits a maximum variation of about 9 \% in the reconstruction errors
of $u_x^{\ast}$, $u_y^{\ast}$ and $T^{\ast}$, 
and the T2F+PINN model shows a slightly larger maximum variation of about 18 \% 
in the reconstruction errors. 
Overall, the T2F and T2F+PINN models demonstrate robust performance across a range of patch sizes
in the RB convection case as well.

\begin{table}
  \centering
  \begin{tabular}{ccccccc}
   & \multicolumn{3}{c}{T2F model} 
   & \multicolumn{3}{c}{T2F+PINN model} \\ 
  Patch size $l_p$ 
   & $u_x^*$ & $u_y^*$ & $T^*$ 
   & $u_x^*$ & $u_y^*$ & $T^*$ \\
  1 & 0.093 & 0.073 & 0.209 & 0.094 & 0.072 & 0.202 \\
  2 & 0.097 & 0.080 & 0.219 & 0.091 & 0.071 & 0.215 \\
  4 & 0.094 & 0.076 & 0.212 & 0.084 & 0.073 & 0.217 \\
  8 & 0.088 & 0.079 & 0.210 & 0.078 & 0.066 & 0.194 \\
  \end{tabular}
  \caption{Normalised $L_2$ errors of the reconstructed velocity components and temperature fields 
  in RB convection for different patch sizes $l_p$. 
  The results are averaged over $n_{\text{test}}=1000$ test samples.}
  \label{tab_lpatch_RB}
\end{table}

Finally, we examine how sensitive the reconstructions are to changes in the control parameters of the RB system. 
Additional RB simulations are performed with Rayleigh numbers spanning $Ra=10^{7}$--$10^{9}$ and aspect ratios ranging from $\Gamma=1$ to 8, while holding $Pr=0.71$ fixed. 
As shown in Appendix~\ref{appRobustness}, a model trained on a single reference case (here $Ra=10^{8},\,\Gamma=2$ with no-slip sidewalls) does not transfer reliably to flows at different $Ra$, $\Gamma$ or wall conditions.
Direct application to new configurations yields large normalised $L_{2}$ errors and often misses the correct large-scale structures. 
In contrast, retraining on the target configuration consistently restores accuracy, typically reducing errors by approximately an order of magnitude. 
These observations indicate that the present models mainly encode trajectory--flow correspondences specific to the training regime, so accurate reconstructions in practice require retraining (or at least fine-tuning) on flows with similar dynamical characteristics.

\section {Conclusion}
\label{sec:conclusion}

In this work, we proposed a deep-learning model, T2F, for reconstructing flow 
fields from sparse, localised trajectories of actively navigating Lagrangian 
agents. The model adopts an encoder--decoder architecture, where a ViT encoder 
captures both local and long-range temporal dependencies in agent motion, and a 
CNN decoder reconstructs the corresponding spatial flow structures. This design 
enables the extraction of rich spatiotemporal representations from limited 
Lagrangian input. To enhance physical fidelity, we further developed a 
physics-informed variant, the T2F+PINN model, by augmenting the data-driven loss 
with equation residuals derived from the governing physical laws. This integration 
of physics-based knowledge into the training process promotes reconstructions that 
are not only data-consistent but also dynamically coherent.

We first validated the model using a laminar cylinder wake flow as 
a proof-of-concept test. The T2F model demonstrated high accuracy in reconstructing 
the velocity field, while the T2F+PINN model significantly improved the 
reconstruction of vorticity. The T2F model outperformed the T2F+PINN model in estimating 
primitive variables due to its purely data-driven optimisation, whereas the 
T2F+PINN model achieved greater accuracy in reconstructing gradient-based 
quantities by incorporating physical constraints. Under varying levels of 
input noise, the T2F+PINN model exhibited enhanced robustness, showing markedly 
lower error growth in vorticity reconstructions even under strong input perturbations.

We then applied the model to the turbulent RB convection, which is a paradigm 
system for convective flow in the atmosphere and oceans. Both the T2F and T2F+PINN 
models accurately reconstructed the primitive variables $u_x^{\ast}$, $u_y^{\ast}$ and $T^{\ast}$, 
but exhibited markedly different performance in gradient-related quantities. 
The T2F+PINN model consistently achieved superior reconstruction accuracy in vorticity, 
temperature gradients and the $Q$ value, outperforming T2F by up to 60.1 \% in 
normalised $L_{2}$ error. These results highlight the capability of the T2F model 
to infer temperature and velocity structures in regions adjacent to sparse Lagrangian 
trajectories, while the T2F+PINN model offers a robust solution for applications 
requiring accurate inference of physically derived quantities.

Beyond demonstrating reconstruction accuracy, our results provide broader 
insights into Lagrangian sensing and data-driven flow reconstruction in turbulent 
environments. We open a promising avenue for real-time, physics-consistent 
inference of flow structures from sparse, localised observations. Owing to its 
data efficiency and robustness, the proposed model is particularly well suited 
for environmental perception tasks in scenarios where global field measurements 
are unavailable, such as soaring flight or underwater navigation. Looking ahead, 
our models can be extended to dynamic flow environments by incorporating 
online-learning strategies that adapt a pre-trained model using only physics-based 
loss functions, thereby eliminating the need for additional labelled data.
It is also worth mentioning that the recently developed novel knowledge-integrated 
additive approach by \cite{zhang2025knowledge} sheds light on the integration 
of physics and machine learning, and may enhance reconstruction by additively 
embedding domain-specific physical constraints directly into our T2F model.

\begin{bmhead}[Supplementary movies.]
Supplementary movies are available at \url{https://doi.org/10.1017/jfm.2025.11033}.
\end{bmhead}

\begin{bmhead}[Acknowledgement.]
The authors acknowledge the Beijing Beilong Super Cloud Computing Co., Ltd for providing HPC resources that have contributed to the research results reported within this paper (\url{http://www.blsc.cn/}).
\end{bmhead}

\begin{bmhead}[Funding.]
This work was supported by the National Natural Science Foundation of China (NSFC) through
  grants nos. 12272311, 12388101, 12125204; the Young Elite Scientists Sponsorship Program by CAST
  (2023QNRC001); and the 111 project of China (project no. B17037).
\end{bmhead}

\begin{bmhead}[Declaration of interests.]
  The authors report no conflict of interest.
\end{bmhead}

\begin{bmhead}[Author ORCIDs.]
Ao Xu, https://orcid.org/0000-0003-0648-2701;	Heng-Dong Xi, https://orcid.org/0000-0002-2999-2694.
\end{bmhead}

\appendix

\section{The T2F model architecture}\label{appArchitecture}

Here, we provide a detailed description of the T2F model architecture,
which is based on a ViT encoder 
and a CNN decoder.

(i) Tokenisation, patch embedding and position embedding.

The ViT processes the input tensor as follows. 
For each instant $t = 1, \dots, l_t$,  the ViT extracts a local spatial patch of size $l_p \times l_p$ 
with $C$ physical variables. 
Flattening this patch yields a vector of length $d_p = l_p^{2}C$, 
referred to as a token.
Collecting tokens across all time steps $l_t$ 
produces the input tensor $x_{\text{token}}\in\mathbb{R}^{l_t \times d_p}$.
This input tensor is then linearly projected into a higher-dimensional
space through a linear layer, a process referred to as patch embedding. 
The transformation is expressed as

\begin{equation}
  x_{\text{embed}} = x_{\text{token}} W_e + b_e,\qquad
  W_e\in\mathbb{R}^{d_p \times d_e},\;\; b_e\in\mathbb{R}^{d_e},
\end{equation}
where $d_e$ is the embedding dimension, 
$W_e$ is a learnable weight matrix and $b_e$ is a learnable bias vector (broadcast and added to each token). 
Finally, a sequence of $l_t$ learnable positional embeddings is added to incorporate temporal positional information, yielding

\begin{equation}
  x_{\text{pos}} = x_{\text{embed}} + p_e,\qquad
  p_e\in\mathbb{R}^{l_t \times d_e},
\end{equation}
where $p_e$ is a learnable positional embedding matrix.

(ii) Vision transformer encoder.

The ViT contains $n_{\text{trans}}$ subencoder layers. 
After patch and positional embedding, the first subencoder layer
takes $x_{\text{pos}}$ as input, 
and its output is subsequently passed to the next subencoder layer. 
Each subencoder layer employs multi-head self-attention (MSA) to extract 
spatiotemporal features from the input $x_{in}$. 

The MSA layers compute the attention scores between all pairs of input tokens. 
The computation is performed across $h$ heads in parallel. 
For each head $h_i$, the input tensor $x_{in}$ is projected into three matrices:

\begin{equation}
  Q_i = x_{in} W_{Q_i}, \quad K_i = x_{in} W_{K_i}, \quad V_i = x_{in} W_{V_i},
\end{equation}
where $W_{Q_i}$, $W_{K_i}$ and $W_{V_i}$ are learnable weight matrices for the query, key and value matrices, respectively. 
The attention scores are then computed as

\begin{equation}
  \text{Attention}(Q_i, K_i, V_i) = \text{softmax}\left(\frac{Q_iK_i^{T}}{\sqrt{d_k}}\right)V_i,
\end{equation}
where $d_k$ is the dimension of the key vectors. 
The outputs of all heads are concatenated and projected back to the original input dimension:

\begin{equation}
  x_{\text{attn}} = \text{concat}(Z_1, Z_2, \ldots, Z_h)W_O,
\end{equation}
where $Z_i = \text{Attention}(Q_i, K_i, V_i)$ and $W_O$ is a learnable weight matrix.

Each MSA layer is followed by a feed-forward network (FFN) with layer normalisation. 
The FFN applies a nonlinear transformation consisting of two linear layers with a ReLU activation function:

\begin{equation}
  x_{\text{FFN}} = \text{ReLU}(x_{\text{attn}}W_\text{FFN1} + b_\text{FFN1})W_\text{FFN2} + b_\text{FFN2},
\end{equation}
where $W_\text{FFN1}$ and $W_\text{FFN2}$ are learnable weight matrices,
and $b_\text{FFN1}$ and $b_\text{FFN2}$ are learnable bias vectors.
The final output of the layer is normalised as 

\begin{equation}
  x_{\text{out}} = \text{LayerNorm}(x_{in} + x_{\text{FFN}}),
\end{equation}
where $\text{LayerNorm}$ normalises each token's feature vector across the 
embedding dimension by computing its mean and variance. 
The final output of the ViT encoder is a tensor, $x_{\text{ViT}}\in\mathbb{R}^{l_t \times d_e}$, 
which represents a latent embedding of the input sequence. 
This latent representation is a learned feature with no explicit physical meaning.
For further details of the ViT, we refer the reader to \cite{dosovitskiy2021vit}.

(iii) Convolutional neural network decoder.

The output $x_{\text{ViT}}$ is linearly projected and 
reshaped to a three-dimensional tensor of shape  $l_{w0} \times l_{w0} \times C_0$,  
where $C_0$ is the number of channels. 
This reshaped tensor is then processed by $n_{\text{CNN}}$ 
transposed convolutional layers, 
each followed by a ReLU activation function. 
These transposed convolutional layers upsample the feature maps
to the desired output resolution. 
For an input $x \in \mathbb{R}^{H_{\text{in}} \times W_{\text{in}} \times C_{\text{in}}}$, 
the transposed convolution operation is defined as

\begin{equation}
  y_{c_{\text{out}}}[i,j] = 
\sum_{c_{\text{in}}=0}^{C_{\text{in}}-1} 
\sum_{m=0}^{H_{\text{in}}-1} 
\sum_{n=0}^{W_{\text{in}}-1} 
x_{c_{\text{in}}}[m,n]
W_{\text{CNN}}[c_{\text{in}}, c_{\text{out}},\, i - ms + p,\, j - ns + p] + 
b_{c_{\text{out}}},
\end{equation}
where $y_{c_{\text{out}}}[i,j]$ is the output feature map at channel $c_{\text{out}}$ and location ($i,j$). 
The kernel $W_{\text{CNN}}$ has shape $C_{\text{in}} \times C_{\text{out}} \times K \times K$, 
with $K$ the kernel size, $s$ the stride and $p$ the padding along height and width.

The final transposed convolutional layer is followed by a 
standard convolutional layer to produce the output tensor. 
For an input $x\in \mathbb{R}^{H_{\text{in}} \times W_{\text{in}} \times C_{\text{in}}}$, 
the convolution operation is given by

\begin{equation}
  y_{c_{\text{out}}}[i,j] =
  \sum_{c_{\text{in}}=0}^{C_{\text{in}}-1}
  \sum_{m=0}^{K-1}
  \sum_{n=0}^{K-1}
  x_{c_{\text{in}}}[\,i s - p + m,\; j s - p + n\,]\,
  W_{\text{CNN}}[c_{\text{out}}, c_{\text{in}}, m, n] + 
  b_{c_{\text{out}}},
\end{equation}
with notations consistent with those of the transposed convolution. 
Further details of these operations are provided in \cite{li2021cnnreview}.

\begin{figure}
  \centering
  \includegraphics[width=0.7\textwidth]{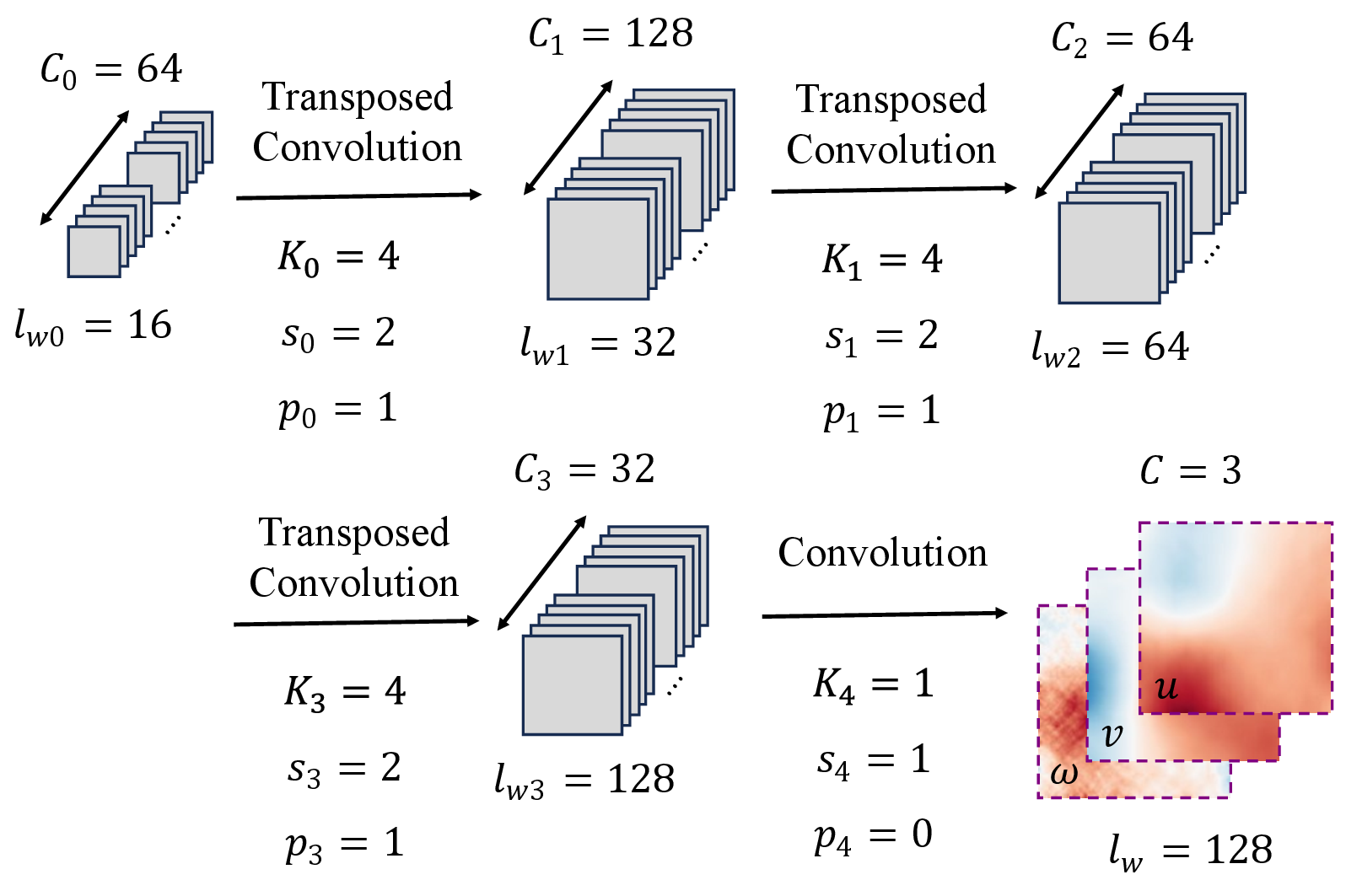}
  \caption{Architecture of the CNN decoder in the T2F model.
  The transposed convolutional layers progressively upsample
  the feature maps to the target resolution $l_w$.}
  \label{fig_t2f_cnn_architecture}
\end{figure}

The final output is a tensor $x_{\text{CNN}}$ of shape $l_{w} \times l_{w} \times C$, 
where $l_{w}$ is the output resolution and $C$ is the number of channels ($C = 2$ for the cylinder wake and $C = 3$ for RB convection). 
The detailed parameters of each layer are shown in figure~\ref{fig_t2f_cnn_architecture}. 

The hyperparameters of the T2F and T2F+PINN models are summarised in table \ref{tab:hyperparameters}, 
including the input trajectory length, patch size, number of transformer layers and attention heads, CNN layers, batch size, learning rate and number of training epochs. 
With these settings, the T2F model contains approximately $2.1\times10^8$ trainable parameters ($\approx$ 811 MB). 
The computational cost for training a single T2F model takes about 4 h on an NVIDIA P100 GPU (16 GB).

\begin{table}
  \centering
  \begin{tabular}{lcc}
    Hyperparameter & Value \\
    Input trajectory length \(l_t\) & 50 \\
    Patch size \(l_p\) & 2 \\
    Number of sub-encoder layers \(n_{\text{trans}}\) & 2 \\
    Embedding dimension \(d_e\) & 256 \\
    Number of attention heads \(h\) & 8 \\
    attention dropout rate & 0.1 \\
    Number of transposed convolutional layers \(n_{\text{CNN}}\) & 3 \\
    Decoder output resolution \(l_{w}\) & 128 \\
    Batch size & 256 \\
    Learning rate & \(1\times10^{-4}\) \\
    Number of training epochs & 10 000 \\
  \end{tabular}
  \caption{Hyperparameters used in the T2F and T2F+PINN models.}
  \label{tab:hyperparameters}
\end{table}

\section{Convergence of the T2F and T2F+PINN models}\label{appConvergence}

\begin{figure}
  \centering
  \includegraphics[width=0.8\textwidth]{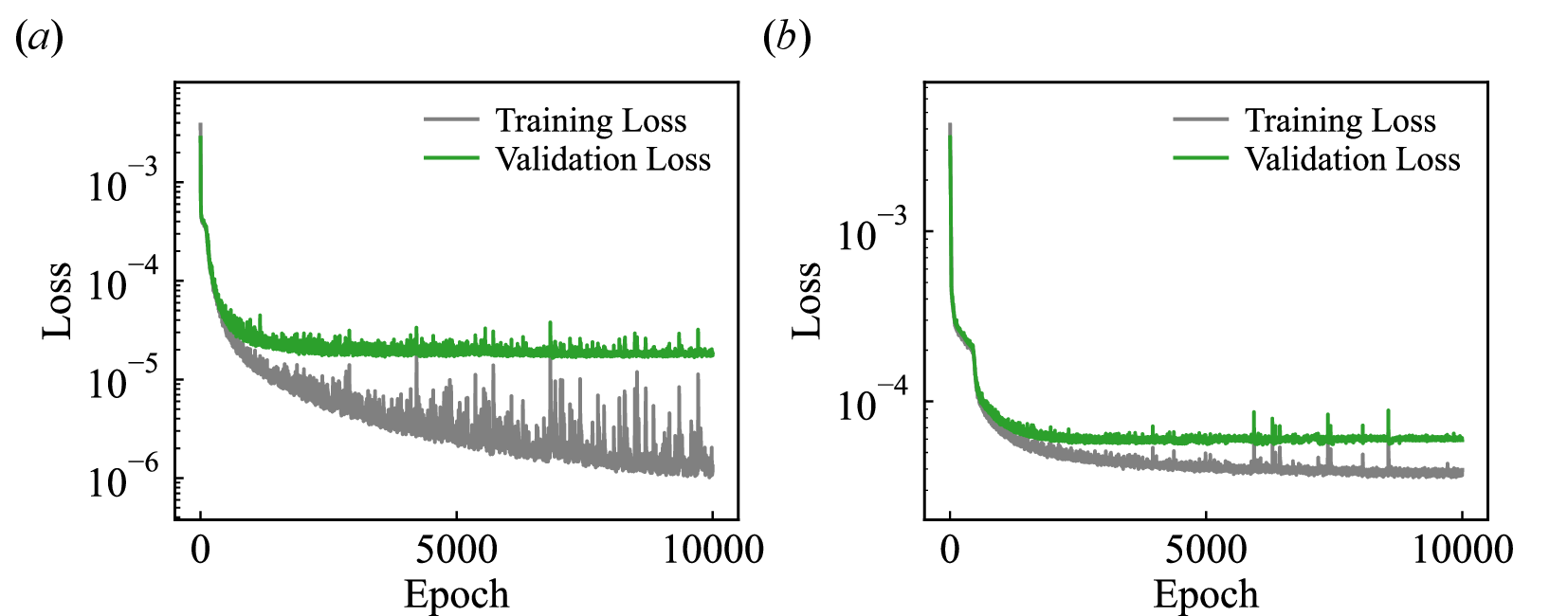}
  \caption{Evolution of the training loss and validation loss for
  (\emph{a}) T2F model and (\emph{b}) T2F+PINN model in the cylinder wake case.}
  \label{fig_appendixC_cylinder}
\end{figure}

\begin{figure}
  \centering
  \includegraphics[width=0.8\textwidth]{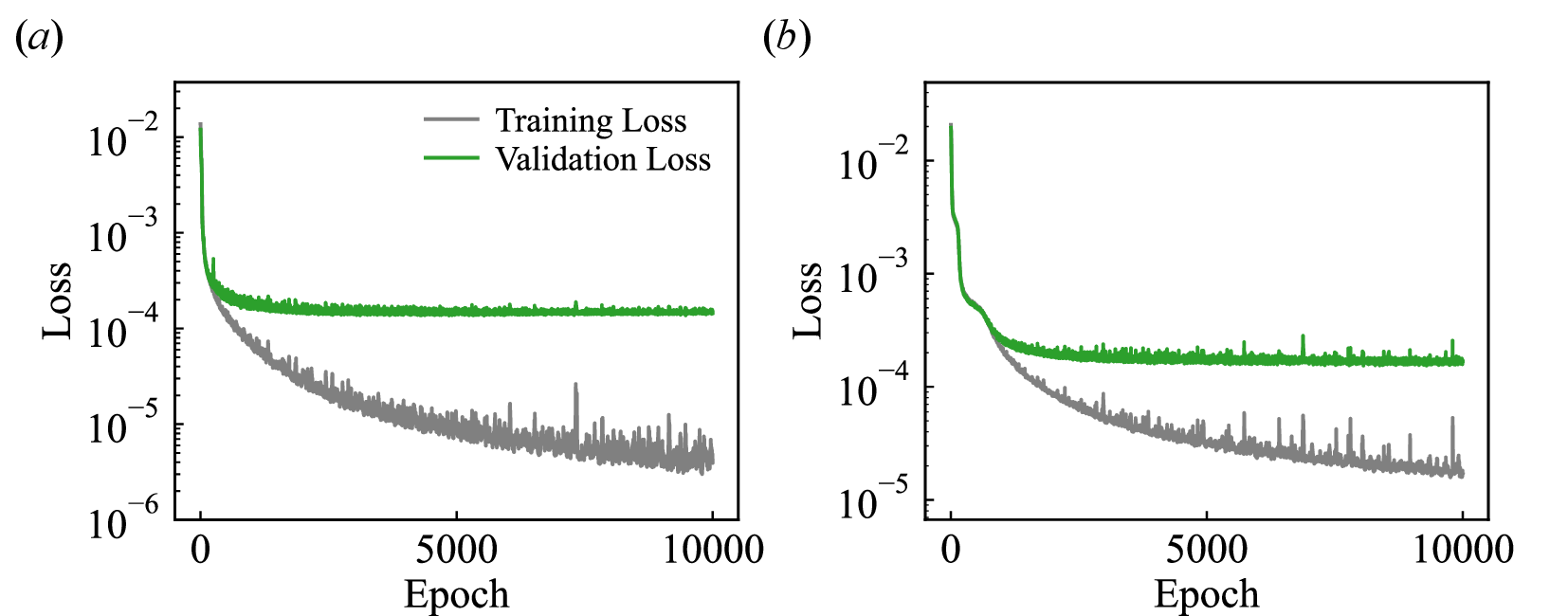}
  \caption{Evolution of the training loss and validation loss for
  (\emph{a}) T2F model and (\emph{b}) T2F+PINN model in the RB convection case.}
  \label{fig_appendixC_RB}
\end{figure}

Here we provide the temporal evolution of both the training and validation losses for the T2F and T2F+PINN models in the cylinder wake case (see figure \ref{fig_appendixC_cylinder}). 
For completeness, the corresponding loss histories for RB 
convection have also been included (see figure \ref{fig_appendixC_RB}). 
Both figures show that, for both models, the validation losses decrease and then plateau after approximately 2000 epochs, indicating that the optimisation reaches a stable minimum. 
This demonstrates that the training of the T2F and T2F+PINN models converges satisfactorily, and that the fluctuations observed in figure \ref{fig_l2error_lineplot}(\textit{a},\textit{b}) reflect variations in test-set performance rather than instability of the training process.

\section{Robustness of the T2F model across different RB convection}\label{appRobustness}

The control parameters in the RB convection system include the Rayleigh number $Ra$, 
the Prandtl number $Pr$ and the aspect ratio $\Gamma$. 
To assess the robustness of the T2F model across different RB convection, 
we conducted additional simulations in which we either apply the same trained T2F model
(see \S~\ref{sec:RB}, trained on $Ra = 10^{8}$, $Pr = 0.71$, $\Gamma = 2$, no-slip sidewalls) 
to different RB configurations (hereafter referred to as `generalization'), 
or retrain the model on each new configuration (hereafter referred to as `retrain'). 
The reconstruction accuracy was assessed using the normalised $L_2$ error of the velocity components $u_x^*$.

Case (i). Similar flow field with plumes rising in the middle.

\begin{figure}
  \centering
  \includegraphics[width=0.75\textwidth]{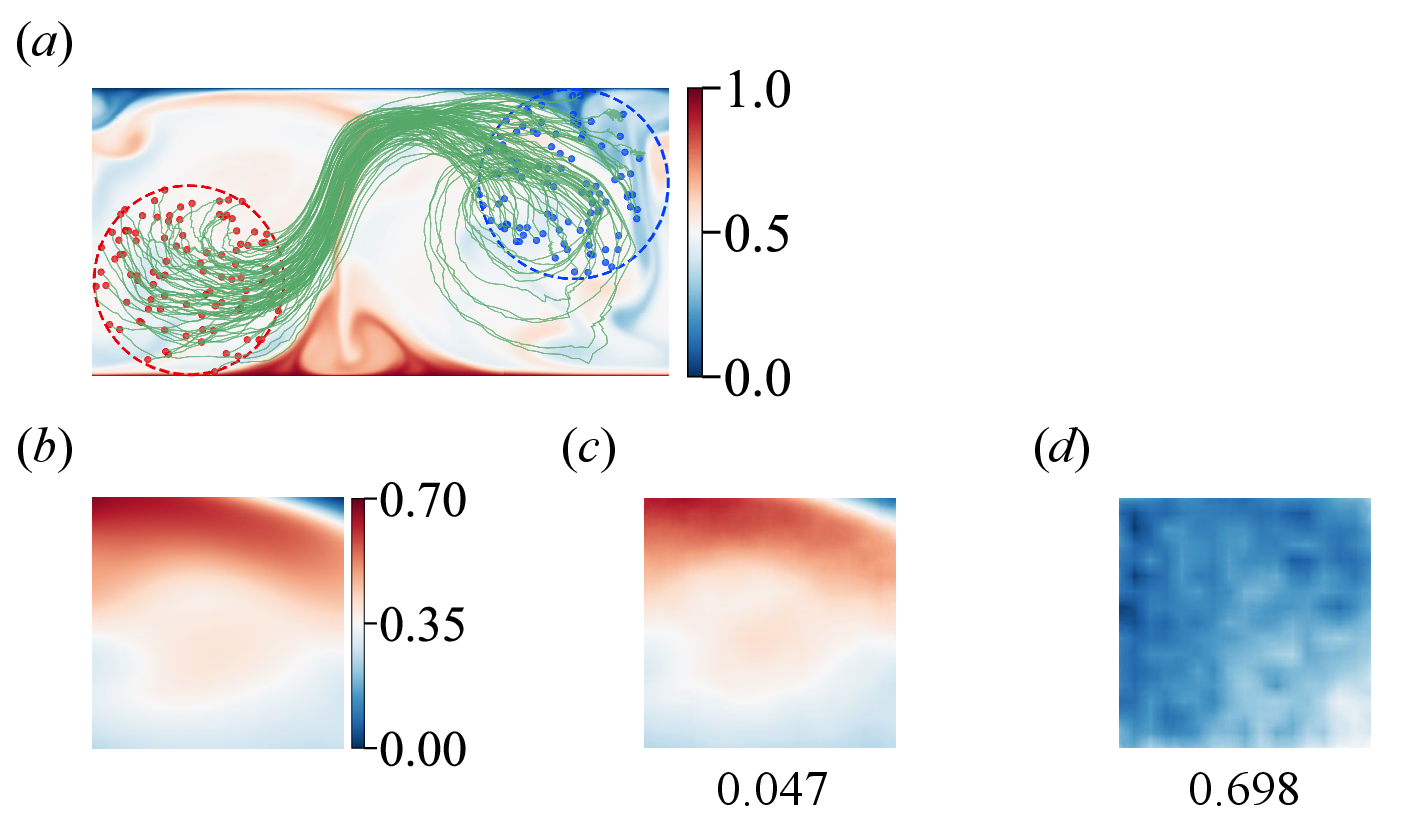}
  \caption{(\emph{a}) Trajectories of self-propelling agents and temperature contours in an RB cell with $\Gamma = 2$ and periodic boundary conditions on the sidewalls.
  (\emph{b}) Ground-truth field of horizontal velocity $u_x^{\ast}$ for a typical sample.
  (\emph{c}) Reconstruction from the retrained model.
  (\emph{d}) Reconstruction from the generalised model.
  Listed values indicate the normalised $L_2$ error.}
  \label{fig_appendixB_gamma2}
\end{figure}

We replaced the no-slip sidewalls with periodic boundary conditions and selected a time period with central plume upwelling for testing. 
Figure \ref{fig_appendixB_gamma2} shows the corresponding trajectories and reconstruction results. 
As summarised in table \ref{tab:fig_appendixB_gamma2}, the generalised model failed to reproduce the correct flow structures, whereas the retrained model achieved accurate reconstructions.

\begin{table}
  \centering
  \begin{tabular}{>{\centering\arraybackslash}p{1cm}
    >{\centering\arraybackslash}p{1cm}
    >{\centering\arraybackslash}p{1.5cm}
    >{\centering\arraybackslash}p{2.2cm}
    >{\centering\arraybackslash}p{1.5cm}}
    \multicolumn{3}{c}{RB convection flow configurations} & \multicolumn{2}{c}{Averaged normalised $L_2$ error} \\
    $Ra$ & $\Gamma$ & Wall type & Generalisation & Retrain \\
    $10^{8}$ & 2 & Periodic & 0.957 & 0.111 \\
  \end{tabular}
  \caption{Averaged reconstruction $L_2$ error over 1000 test samples 
  for generalisation and retraining in an RB cell with $Ra=10^{8}$, 
  $\Gamma=2$ and periodic boundary condition for sidewalls.}
  \label{tab:fig_appendixB_gamma2}
\end{table}

Case (ii). Flow transition from double rolls to a single large roll.

\begin{figure}
  \centering
  \includegraphics[width=0.75\textwidth]{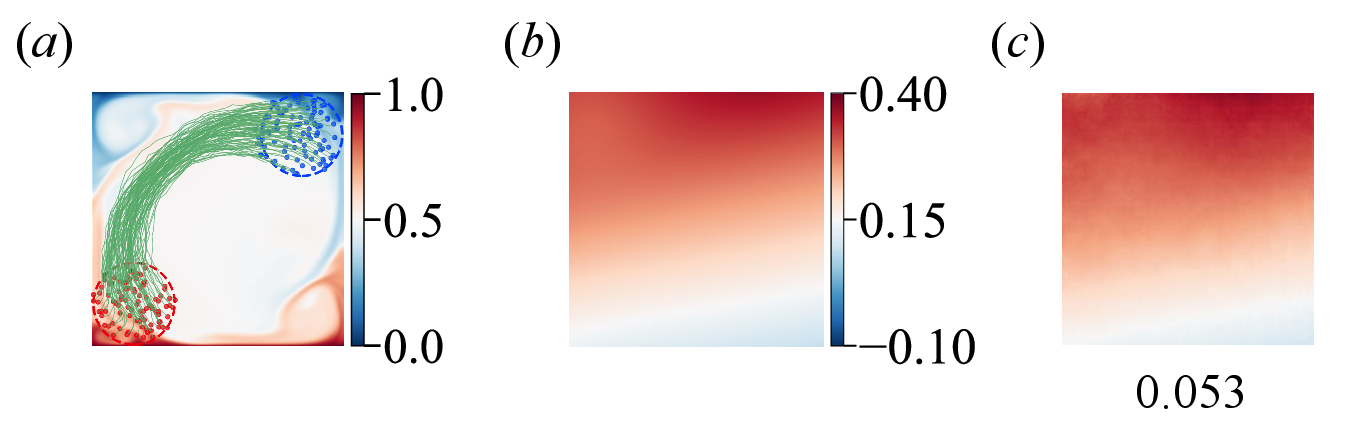}
  \caption{(\emph{a}) Trajectories of self-propelling agents and temperature contours in an RB cell with $\Gamma = 1$.
  (\emph{b}) Ground-truth field of horizontal velocity $u_x^{\ast}$ for a typical sample.
  (\emph{c}) Reconstruction from the retrained model.
  Listed values indicate the normalised $L_2$ error.}
  \label{fig_appendixB_gamma1}
\end{figure}

We set the aspect ratio $\Gamma=1$, where the flow exhibits a single-roll state. 
Due to the reduced domain size, only shorter trajectories were available, making generalisation infeasible. As shown in figure \ref{fig_appendixB_gamma1} and table \ref{tab:fig_appendixB_gamma1}, reliable reconstruction was achieved after retraining.

\begin{table}
  \centering
  \begin{tabular}{>{\centering\arraybackslash}p{1cm}
    >{\centering\arraybackslash}p{1cm}
    >{\centering\arraybackslash}p{2.5cm}
    >{\centering\arraybackslash}p{2.2cm}
    >{\centering\arraybackslash}p{1.5cm}}
    \multicolumn{3}{c}{RB convection flow configurations} & \multicolumn{2}{c}{Averaged normalised $L_2$ error} \\
    $Ra$ & $\Gamma$ & Wall type & Generalisation & Retrain \\
    $10^{8}$ & 1 & No-slip wall & -- & 0.056 \\
  \end{tabular}
  \caption{Averaged reconstruction $L_2$ error over 1000 test samples for generalisation and retraining in an RB cell with 
  $Ra=10^{8}$ and $\Gamma=1$.}
  \label{tab:fig_appendixB_gamma1}
\end{table}

Case (iii). Different aspect ratios and/or Rayleigh numbers. 

\begin{figure}
  \centering
  \includegraphics[width=0.75\textwidth]{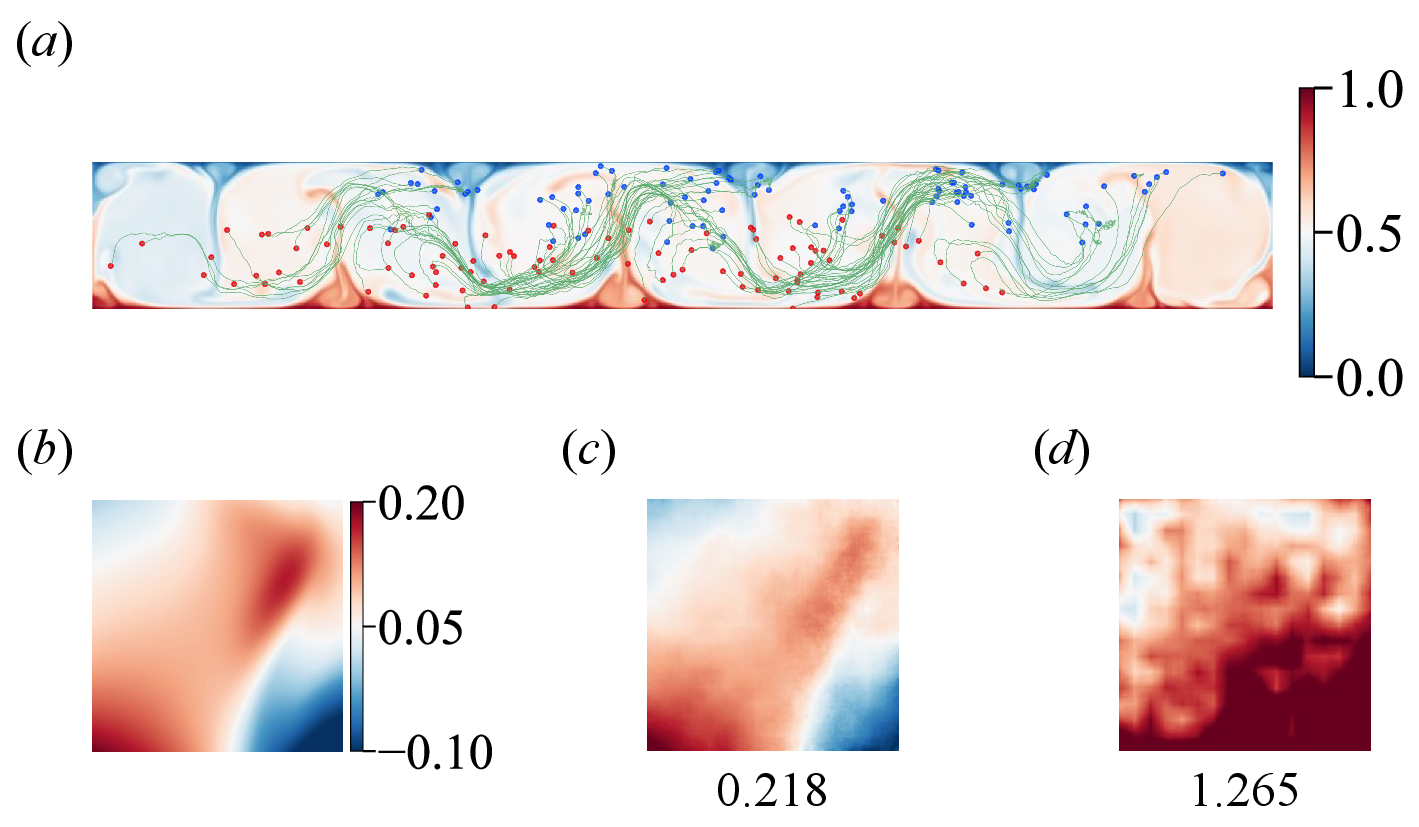}
  \caption{(\emph{a}) Trajectories of self-propelling agents and temperature contours in an RB cell with
  $Ra = 10^{8}$ and $\Gamma = 8$.
  (\emph{b}) Ground-truth field of horizontal velocity $u_x^{\ast}$ for a typical sample.
  (\emph{c}) Reconstruction from the retrained model.
  (\emph{d}) Reconstruction from the generalised model.
  Listed values indicate the normalised $L_2$ error.}
  \label{fig_appendixB_gamma8}
\end{figure}

\begin{figure}
  \centering
  \includegraphics[width=0.75\textwidth]{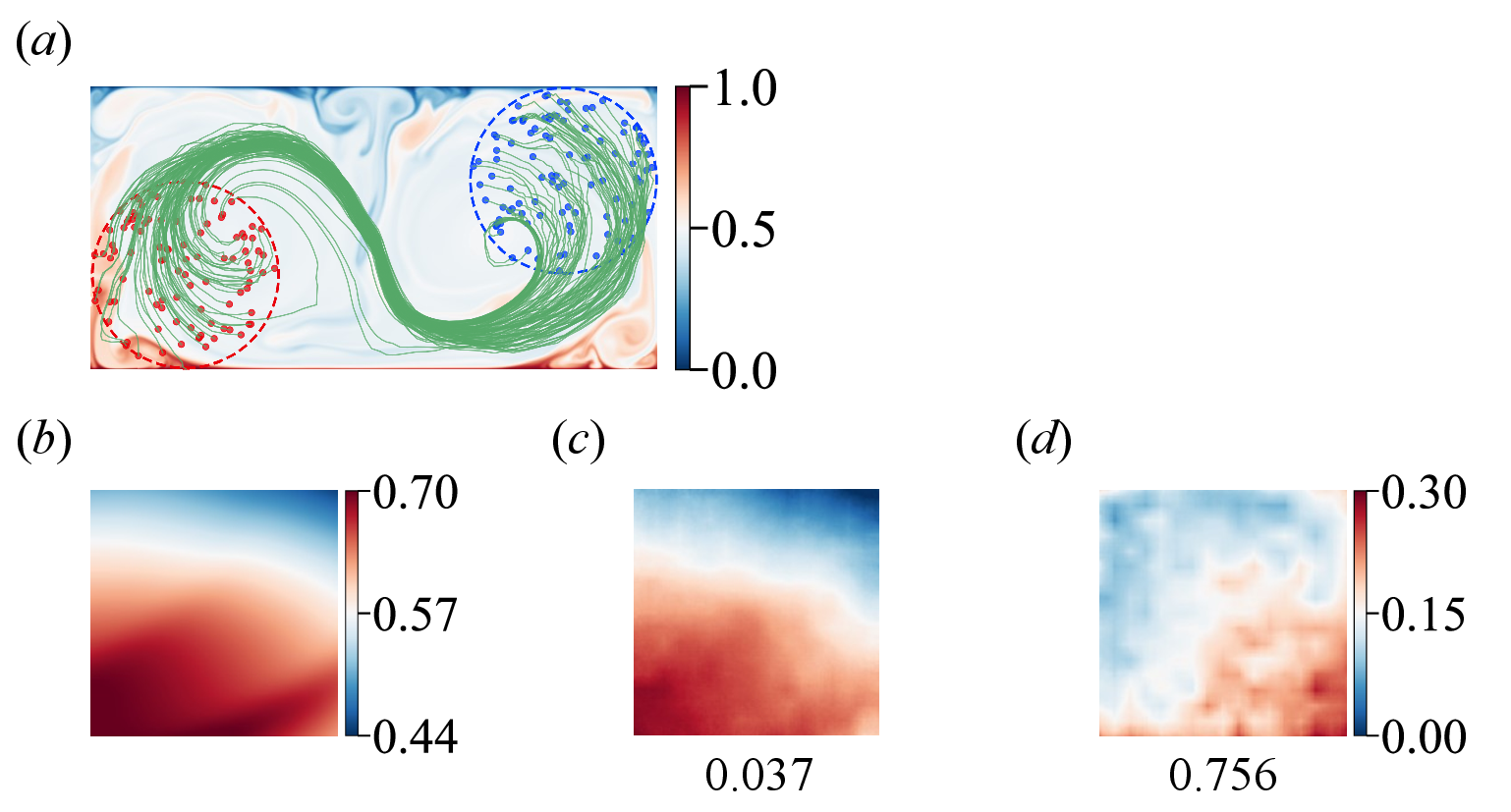}
  \caption{(\emph{a}) Trajectories of self-propelling agents and temperature contours in an RB cell with
  $Ra = 10^{9}$ and $\Gamma = 2$.
  (\emph{b}) Ground-truth field of horizontal velocity $u_x^{\ast}$ for a typical sample.
  (\emph{c}) Reconstruction from the retrained model.
  (\emph{d}) Reconstruction from the generalised model.
  Listed values indicate the normalised $L_2$ error.}
  \label{fig_appendixB_Ra1e9}
\end{figure}

We tested cases with aspect ratio $\Gamma = 4-8$ and Rayleigh number $Ra$ ranging from $10^7$ to $10^9$ (see figures \ref{fig_appendixB_gamma8} and \ref{fig_appendixB_Ra1e9} and table \ref{tab:fig_appendixB_diff_Ra}). 
Across these conditions, generalisation errors remained large, while retrained models consistently reduced errors by an order of magnitude.

\begin{table}
  \centering
  \begin{tabular}{>{\centering\arraybackslash}p{1cm}
    >{\centering\arraybackslash}p{1cm}
    >{\centering\arraybackslash}p{2.5cm}
    >{\centering\arraybackslash}p{2.2cm}
    >{\centering\arraybackslash}p{1.5cm}}
    \multicolumn{3}{c}{RB convection flow configurations} & \multicolumn{2}{c}{Averaged normalised $L_2$ error} \\
    $Ra$ & $\Gamma$ & Wall type & Generalisation & Retrain \\
    $10^{8}$      & 4 & No-slip wall & 1.132 & 0.161 \\
    $10^{8}$      & 8 & No-slip wall & 1.008 & 0.143 \\
    $10^{7}$      & 2 & No-slip wall & 1.628 & 0.064 \\
    $2\times10^{7}$ & 2 & No-slip wall & 1.850 & 0.182 \\
    $5\times10^{7}$ & 2 & No-slip wall & 1.807 & 0.138 \\
    $2\times10^{8}$ & 2 & No-slip wall & 1.021 & 0.100 \\
    $5\times10^{8}$ & 2 & No-slip wall & 0.898 & 0.116 \\
    $10^{9}$      & 2 & No-slip wall & 1.138 & 0.115 \\
  \end{tabular}
  \caption{Averaged reconstruction $L_2$ errors over 1000 test 
  samples for generalisation and retraining in RB cells with 
  different aspect ratio $\Gamma$ and Rayleigh number $Ra$.}
  \label{tab:fig_appendixB_diff_Ra}
\end{table}


\bibliographystyle{jfm}

\end{document}